\documentclass[conference,compsoc]{IEEEtran}

\ifCLASSOPTIONcompsoc
  \usepackage{cite}
\else
  \usepackage{cite}
\fi

\ifCLASSINFOpdf
\else
\fi

\usepackage{tikz}
\usepackage{titlesec}
\usepackage{ulem}  
\usepackage{paralist}

\newcommand{\subject}[1]{\vspace{3pt}\noindent\textbf{#1}}

\newcommand{\newtext}[1]{\textcolor{black}{#1}}

\usepackage[dvipsnames]{xcolor} 
\usepackage{xspace}
\usepackage{graphicx}
\usepackage{enumitem}
\usepackage{url}
\usepackage{hyperref}
\usepackage{amsmath}
\usepackage{amssymb}
\usepackage{float}

\usepackage{caption}
\usepackage{subcaption}
\usepackage{lipsum}
\usepackage{tabularray}
\usepackage{tcolorbox}
\usepackage{booktabs}
\usepackage{newtxmath}
\usepackage{newtxtext}
\usepackage{newtxtt}
\usepackage{array}
\usepackage{multirow}
\usepackage{subcaption}

\usepackage{algorithm}
\usepackage{algpseudocode}
\usepackage{eurosym}
\usepackage{pifont}
\newcommand{\tool}{\textit{Cosmic}\xspace}

\begin{document}

\newcommand{\pone}{%
  \begin{tabular}[t]{@{}l@{}}
    \textcolor{blue}{$\mathcal{P}_1$}(R, P, E) $\Leftarrow$ \\
    \quad \texttt{\textcolor{RedViolet}{Action}}(R, A), \\
    \quad \texttt{\textcolor{RedViolet}{Purpose}}(R, P), \\
    \quad \texttt{\textcolor{RedViolet}{ActionMapping}}(E, A), \\
    \quad \texttt{\textcolor{RedViolet}{Item}}(E,\_,\_).
  \end{tabular}%
}

\newcommand{\ptwo}{%
  \begin{tabular}[t]{@{}l@{}}
    \textcolor{blue}{$\mathcal{P}_2$}(R, P, E) $\Leftarrow$ \\
    \quad \texttt{\textcolor{RedViolet}{Purpose}}(R, P), \\
    \quad \texttt{\textcolor{RedViolet}{ElementSent}}(E, R), \\
    \quad \texttt{\textcolor{RedViolet}{Item}}(E, E\_type, \_), \\
    \quad \texttt{\textcolor{RedViolet}{IsInSelectType}}(E\_type).
  \end{tabular}%
}

\newcommand{\pthree}{%
  \begin{tabular}[t]{@{}l@{}}
    \textcolor{blue}{$\mathcal{P}_3$}(P,E) $\Leftarrow$ \\
    \quad (\textcolor{blue}{$\mathcal{P}_1$}(\_, P, \_) ; \textcolor{blue}{$\mathcal{P}_2$}(\_, P, \_)), \\
    \quad \texttt{\textcolor{RedViolet}{Category}}(\_, C, P), \\
    \quad count : \{ C : (\textcolor{blue}{$\mathcal{P}_1$}(\_, P, \_) ; \textcolor{blue}{$\mathcal{P}_2$}(\_, P, \_)), \\
    \quad \texttt{\textcolor{RedViolet}{Category}}(\_, C, P), \\
    \quad \texttt{\textcolor{RedViolet}{Item}}(E, \_, \_) \} = 1, \\
    \quad \texttt{\textcolor{RedViolet}{SubmitButton}}(E)
  \end{tabular}%
}

\newcommand{\pfour}{%
  \begin{tabular}[t]{@{}l@{}}
    \textcolor{blue}{$\mathcal{P}_4$}(P, E) $\Leftarrow$ \\
    \quad (\textcolor{blue}{$\mathcal{P}_1$}(R, P, E) ; \textcolor{blue}{$\mathcal{P}_2$}(R, P, E)), \\
    \quad \texttt{\textcolor{RedViolet}{Category}}(\_, C, P), \\
    \quad count : \{ C : (\textcolor{blue}{$\mathcal{P}_1$}(\_, P, E) ; \textcolor{blue}{$\mathcal{P}_2$}(\_, P, E)), \\
    \quad \texttt{\textcolor{RedViolet}{Category}}(\_, C, P), \\
    \quad \texttt{\textcolor{RedViolet}{Item}}(E, \_, \_) \} = 1.
  \end{tabular}%
}


\newcommand{\pfive}{%
  \begin{tabular}[t]{@{}l@{}}
    \textcolor{blue}{$\mathcal{P}_5$}(True) $\Leftarrow$ \\
    \quad \texttt{\textcolor{RedViolet}{Withdraw}}(\_, \_)
  \end{tabular}%
}

\newcommand{\psix}{%
    \begin{tabular}[t]{@{}l@{}}
    $\begin{array}{@{}l@{}}
      \textcolor{blue}{\mathcal{P}_6}(C) \Leftarrow \\
      \quad \textcolor{RedViolet}{\texttt{Controller}}(C), C \neq null \\
    \end{array}$%
    \end{tabular}%
}

\newcommand{\pseven}{%
  \begin{tabular}[t]{@{}l@{}}
    \textcolor{blue}{$\mathcal{P}_7$}(P) $\Leftarrow$ \\
    \quad \texttt{\textcolor{RedViolet}{Purpose}}(P), $P \neq \text{null}$ \\
  \end{tabular}%
}

\newcommand{\peight}{%
  \raisebox{-0.4\height}{$\begin{array}{@{}l@{}}
    \textcolor{blue}{\mathcal{P}_8}(E) \Leftarrow \\
    \quad \textcolor{blue}{\mathcal{P}_2}(\_, \_, E), \neg \textcolor{RedViolet}{\texttt{Selected}}(E) \\
  \end{array}$}%
}

\newcommand{\pnine}{%
  \raisebox{-0.5\height}{$\begin{array}{@{}l@{}}
    \textcolor{blue}{\mathcal{P}_9}(E) \Leftarrow \\
    \quad (\textcolor{blue}{\mathcal{P}_1}(R, \_, E); \textcolor{blue}{\mathcal{P}_2}(R, \_, E)),\\
    \quad \textcolor{RedViolet}{\texttt{Polarity}}(R),\\
    \quad \textcolor{RedViolet}{\texttt{Item}}(E, \_, \_). \\
  \end{array}$}%
}

\title{Breaking the Illusion: Automated Reasoning of GDPR Consent Violations}
\author{
    \IEEEauthorblockN{
        Ying Li\IEEEauthorrefmark{1},
        Wenjun Qiu\IEEEauthorrefmark{2},
        Faysal Hossain Shezan\IEEEauthorrefmark{3},
        Kunlin Cai\IEEEauthorrefmark{1},
        Michelangelo van Dam\IEEEauthorrefmark{4},\\
        Lisa Austin\IEEEauthorrefmark{2},
        David Lie\IEEEauthorrefmark{2}, and
        Yuan Tian\IEEEauthorrefmark{1}
    }
    \IEEEauthorblockA{
        \IEEEauthorrefmark{1} University of California, Los Angeles \IEEEauthorrefmark{2} University of Toronto
        \IEEEauthorrefmark{3} University of Texas at Arlington
        \IEEEauthorrefmark{4} in2it \\
        \{yinglee, kunlin96, yuant\}@ucla.edu, \{wenjun.qiu, lisa.austin, david.lie\}@utoronto.ca, \\ faysal.shezan@uta.edu, michelangelo@in2it.be
    }
}
\maketitle

\begin{abstract}
Recent privacy regulations such as the General Data Protection Regulation (GDPR) and the California Consumer Privacy Act (CCPA) have established legal requirements for obtaining user consent regarding the collection, use, and sharing of personal data. These regulations emphasize that consent must be informed, freely given, specific, and unambiguous. However, there are still many violations, which highlight a gap between legal expectations and actual implementation.
Consent mechanisms embedded in functional web forms across websites play a critical role in ensuring compliance with data protection regulations such as the GDPR and CCPA, as well as in upholding user autonomy and trust. However, current research has primarily focused on cookie banners and mobile app dialogs.
These forms are diverse in structure, vary in legal basis, and are often difficult to locate or evaluate, creating a significant challenge for automated consent compliance auditing. 
In this work, we present \tool, a novel automated framework for detecting consent-related privacy violations in web forms. 
\tool integrates three key innovations: 
(1) a large language model (LLM)-based framework to extract consent requirements from privacy policies and locate relevant forms using multimodal web agents; 
(2) a domain-specific language (DSL) to formally describe heterogeneous web form structures, enabling systematic analysis; and 
(3) machine-interpretable Datalog rules, derived in collaboration with privacy experts, to translate natural-language GDPR requirements into formal logic for automated verification. 
We evaluate our developed tool for auditing consent compliance in web forms, across 5,823 websites and 3,598 forms. 
\tool detects 3,384 violations on 94.1\% of consent forms, covering key GDPR principles such as freely given consent, purpose disclosure, and withdrawal options. 
It achieves 98.6\% and 99.1\% TPR for consent and violation detection, respectively, demonstrating high accuracy and real-world applicability.
\end{abstract}

\IEEEpeerreviewmaketitle

\section{Introduction}

Privacy regulations such as the General Data Protection Regulation (GDPR)~\cite{gdprconsent} and the California Consumer Privacy Act (CCPA)~\cite{ccpa} emphasize individuals' control over their personal data, with consent as a foundational principle. 
These regulations require organizations to obtain explicit, informed, freely given, and unambiguous consent before collecting or processing personal information. 
Enforcing these requirements is essential to prevent deceptive data practices and protect user privacy in increasingly complex digital environments. 
However, widespread non-compliance persists, and regulatory bodies have responded with significant enforcement actions. 
By January 2025, cumulative GDPR fines had reached approximately \euro 5.88 billion, with consent violations accounting for 30\% of the 20 largest penalties~\cite{gdpr20fines}. 
Notably, Amazon was fined \euro 746 million in 2021 for operating its advertising system without proper consent~\cite{amazon-gdpr-fine}.



Despite these high-profile enforcement actions and significant financial penalties, consent compliance remains challenging to enforce across different interfaces. In particular, websites are central to personal data collection through forms, trackers, and interactive services and thus play a critical role in upholding user consent requirements. Prior research for web consent has focused on analyzing cookie consent banners~\cite{bollinger2022automating, matte2020cookie, bouhoula2024automated, bielova2024effect, papadogiannakis2021user, bui2022opt, klein2022accept, utz2019informed}. 

\begin{figure*}[t!]
    \centering
    \begin{subfigure}{0.24\textwidth}
        \includegraphics[width=\textwidth, keepaspectratio]{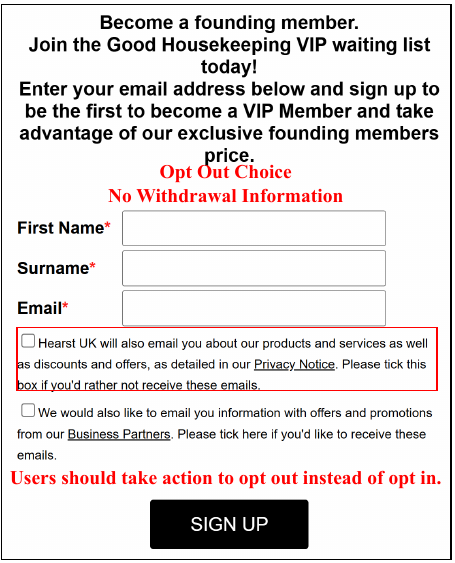}
        \caption{}
        \label{fig:scenarios_0}
    \end{subfigure}\hfill
    \begin{subfigure}{0.24\textwidth}
        \includegraphics[width=\textwidth,  keepaspectratio]{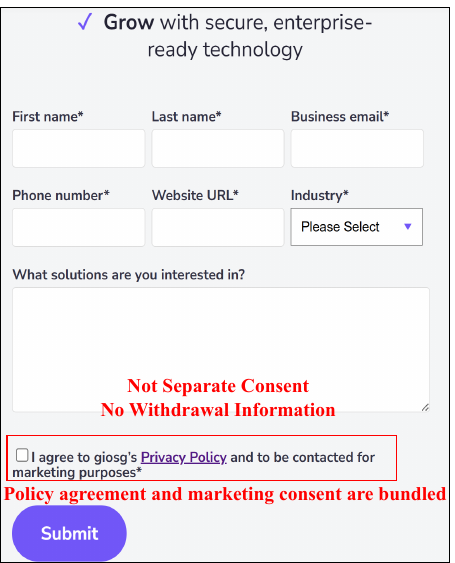}
        \caption{}
        \label{fig:scenarios_1}
    \end{subfigure}\hfill
    \begin{subfigure}{0.24\textwidth}
        \includegraphics[width=\textwidth, keepaspectratio]{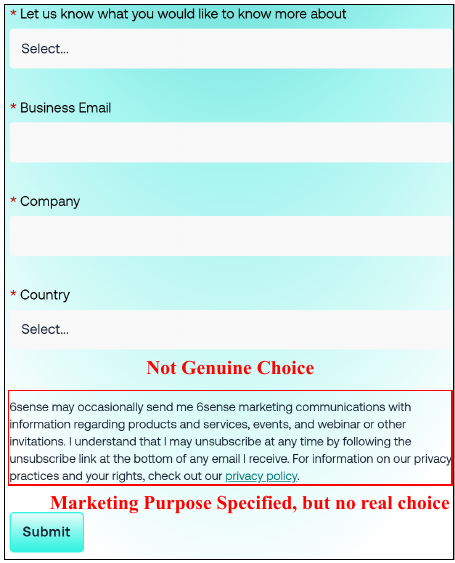}
        \caption{}
        \label{fig:scenarios_2}
    \end{subfigure}\hfill
    \begin{subfigure}{0.24\textwidth}
        \includegraphics[width=\textwidth, keepaspectratio]{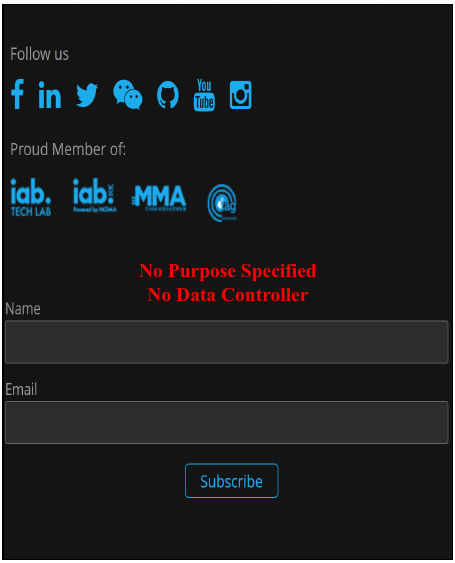}
        \caption{}
        \label{fig:scenarios_3}
    \end{subfigure}
    \caption{Four different types of privacy violations in consent mechanisms on web forms from real world.}
    \label{fig:example_all} 
\end{figure*}

However, a critical yet underexplored vector is the consent compliance of web forms: key mechanisms for collecting personal information online. While cookie banners ask permission to collect tracking information like device fingerprint and preferences, web forms collect more identifiable data like users' age, name, and date of birth directly from user input~\cite{cui2025webform}. 
Consequently, this research gap puts users at risk when they consent to web form data collection without understanding the real implications. For example, poorly designed forms can even mislead users into giving unintended consent--for instance, booking a demo may inadvertently opt them into marketing communications (Figure~\ref{fig:scenarios_2}). 
In addition, unlike cookie banners, web forms vary widely in design and placement, are complex structurally, and discovering web forms requires interactive browsing and engagement with websites. 

Therefore, studying the compliance of 
consent web forms remains an open and challenging problem.
\textbf{First}, a significant challenge lies in bridging the gap between real-world implementations and the textually defined requirements of GDPR. 
Even minor discrepancies can lead to unintentional privacy violations, and the natural language nature of GDPR makes compliance verification especially difficult without expert interpretation. Therefore, a formal way to represent the forms and check violations is essential to provide formal guarantees of compliance accuracy, and consistency.
\textbf{Second}, GDPR Article 6~\cite{GDPR_Art6} defines multiple legal basis for data processing, meaning not all forms require user consent. 
Thus, we must analyze privacy policies to identify the claimed legal basis (i.e., when consent is required) before evaluating consent interfaces.
However, this is difficult because privacy policies are unstructured, often spanning multiple web pages, which complicates the identification of specific data processing purposes (e.g., subscriptions) that necessitate consent. In addition, it is challenging to match the legal basis in the privacy policies with the corresponding consent forms. 
\textbf{Third}, the diverse and dynamic nature of websites hinders the automatic detection of web forms that collect personal data. 
Unlike cookie banners, which typically appear at page load, consent forms can be embedded anywhere--as pop-ups, modal dialogs, or within functional pages--making them harder to locate.
\textbf{Finally},  the structural and presentational variability of consent forms further complicates scalable identification of compliance violations across websites.

To address the aforementioned challenges, 
we propose a novel automated reasoning framework called \tool (\underline{\textbf{Co}}n\underline{\textbf{s}}ent co\underline{\textbf{m}}pl\underline{\textbf{i}}ance \underline{\textbf{C}}hecker).
The key insights of \tool are: 
(1) We bridge the gap between natural language GDPR requirements and automated verification by working with privacy law experts to rigorously convert these requirements into machine-interpretable Datalog rules. We then leverage off-the-shelf verification technologies to automatically detect privacy violations through formal reasoning. \tool generates a violation report that identifies specific violations, their locations, and the corresponding GDPR requirements that are violated.
(2) We develop a Domain-Specific Language (DSL) to represent the consent-related implementation in web forms.
This DSL enables scalable formal verification and can be used to automatically detect consent violations in other applications beyond the initial web form context.
(3) To address the challenge of understanding the legal basis in privacy policies, we employ a Retrieval-Augmented Generation (RAG)-based approach to extract consent-related passages (could be a sentence, or a paragraph), and determine when consent is required through LLM-based in-context learning.
(4) To address the challenge of matching legal basis in privacy policies with specific forms and analyzing the complex and interactive web forms, we introduce an agentic approach that automatically and dynamically locates corresponding consent forms across diverse website structures. Through the information extracted in the screenshot, HTML, and accessibility Tree, we propose a fuzzy match algorithm to ensure accurate form extraction. 

Putting these together, our evaluation demonstrates that \tool effectively detects consent compliance violations with 98.6\% TPR and 99.1\% TPR. We applied \tool to analyze 5,823 websites from 70+ categories with 3,598 web forms, discovering 3,384 consent compliance violations across 94.1\% of analyzed consent forms. These violations span various GDPR principles, including failures to ensure freely given consent, lack of specific purpose disclosure, missing withdrawal information, and inappropriate use of pre-selected consent options. 

\subject{Contributions.} We summarize our contributions as follows:
\begin{itemize}
    \item We conducted the first comprehensive study of GDPR consent compliance violations in website consent forms, particularly focusing on consent forms and their compliance with GDPR consent requirements, revealing widespread non-compliance across diverse website categories.
    \item We design and implement \tool\footnote{The tool is available at \url{https://github.com/UCLA-Security-and-Privacy-Lab/Cosmic}}, a tool that can automatically detect GDPR consent violations through formal reasoning. \tool is based on our new DSL for modeling consent forms and machine-interpretable encoding of GDPR requirements as Datalog rules.
    \item We explored new insights into the prevalence of consent violations across diverse website categories, revealing that 94.1\% of analyzed consent forms contain violations, and discussed the implications for improving consent mechanism design and enhancing user privacy protection in web applications.
\end{itemize}


\subject{Roadmap.} We first provide background in Section~\ref{sec:background}. Then, we present the motivating examples and challenges in Section~\ref{sec:example}. Next, we show \tool's design and evaluation in Section~\ref{sec:method} and~\ref{sec:experiment}. We discuss limitations and related work in Section~\ref{sec:discussion} and~\ref{sec:related_work}, and conclude in Section~\ref{sec:conclusion}.

\section{Background}
\label{sec:background}

\subsection{Legal Basis of GDPR}

To ensure lawful data processing, GDPR requires organizations to identify a valid legal basis -- a legitimate reason for collecting and using personal data. This choice is fundamental to privacy compliance as it determines both user rights and organizational obligations. GDPR~\cite{GDPR_Art6} outlines six legal basis under which personal data may be lawfully processed: consent, contractual necessity, legal obligation, vital interests, public task, and legitimate interests.

However, \textit{Consent won't always be the most appropriate or easiest} legal basis, as emphasized by UK ICO~\cite{ico_consent_appropriate}. 
Controllers must choose the legal basis that most accurately reflects the true nature of their relationship with the individual and the specific purpose of processing. 

Nevertheless, in certain situations, consent is not only appropriate but legally required.
For example, under the UK Privacy and Electronic Communications Regulations (PECR)~\cite{PECR,PECR_2003_2426}, consent is mandatory for marketing calls, emails, texts, and faxes, etc., regardless of whether another legal basis could justify the processing under GDPR. In such cases, consent becomes the sole lawful mechanism to legitimize the data collection activity. 

In the context of consent mechanisms in websites, \textit{legal obligation} (Art. 6(1)(c) GDPR~\cite{GDPR_Art6}) does not apply, as such data is not required to comply with a legal duty; \textit{vital interests}(Art. 6(1)(d)) are limited to situations concerning the protection of life or physical integrity, which do not arise in routine web interactions. These basis are therefore not relevant for the types of optional data collection commonly performed through website forms. While legitimate interest (Art. 6(1)(f)) may apply to certain website activities such as basic analytics and security measures, it cannot be relied upon for marketing communications, etc, which typically require explicit consent under GDPR and ePrivacy rules.  The determination of appropriate legal basis is highly context-dependent and requires a case-by-case assessment. Even identical web forms may have different legal requirements depending on how the privacy policy frames the data processing purposes and the specific implementation context.

In practice, websites often disclose their intended legal basis in privacy policies, mapping different processing purposes to specific basis—e.g., contractual necessity for registration, and consent for tracking or advertising. However, regardless of these claims, compliance depends not only on the correctness of the declaration but also on how the corresponding mechanism is implemented at the user interface level. In particular, when consent is declared as the legal basis, the interface must provide a valid mechanism for obtaining it under Article 7 of the GDPR.

\subsection{Datalog}
Logic programming languages are powerful tools for modeling and reasoning about complex systems using well-defined rules. It is well-suited for applications where accuracy and transparency are critical~\cite{kim2025large}. In our work, we use Datalog as the Logic Programming Language. 

Datalog~\cite{wikipedia_datalog_2025} is a simplified version of general \textit{Logic Programming}~\cite{lloyd2012foundations}.
A Datalog program consists of a finite set of \textit{rules} and \textit{facts}. 
Facts are assertions about the world, for example, \textit{mother("Alice", "Bob")} states that Alice is Bob's mother. Rules allow us to deduce facts from existing ones. Consider the rule: ``if $X$ is $Y$'s mother and if $Y$ is $Z$'s mother, then $X$ is a grandma of $Z$''. Rules typically contain variables ($X$, $Y$, $Z$ in this example) to express general patterns of inference.
Formally, both Datalog facts and rules are represented as \textit{Horn clauses} with the syntax: $L_0: L_1,\cdots, L_n$, where each $L_i$ is a literal of the form $p_i(t_1,\cdots,t_{k_i})$, $p_i$ is \textit{predicate} symbol and $t_j$ are terms. A term is either a \textit{constant}(e.g., "Alice") or \textit{variable} (e.g., $X$).
The left-hand side of a Datalog clause is called \textit{head} and the right hand side is called \textit{body} (e.g.,\textit{mother("Alice", "Bob")} ), while clauses with non-empty bodies represent rules that specify conditions under which the head can be inferred.

In our work, we leverage Datalog's formal reasoning capabilities to automatically detect GDPR consent violations by encoding consent form properties as facts and GDPR requirements as rules.

\section{Motivating Example}
\label{sec:example}



In this section, we introduce examples of GDPR consent violations from real-world websites as motivation of this research.
Consider a user named Alice who browses the web and interacts with different privacy consent mechanisms.
Alice may experience problematic consent implementations.



In scenario 1, Alice visits \textit{\url{www.goodhousekeeping.com}} and becomes interested in joining their VIP waiting list. 
Upon clicking `Subscribe', she is promoted to the form in Figure~\ref{fig:scenarios_0}. 
Alice doesn't want promotional emails, so it is natural for her to assume that she leaving any checkbox unmarked aligns with her preference.
However, the form's instructions are confusing and misleading.
Rather than requiring checking the box for agreement to receive emails, the prompt instead indicates a desire to opt out.
This inverted logic will lead Alice to unintentionally consent to receive communications she wish to avoid,
violating GDPR Article 7~\cite{GDPR_Art7}.
The artifact mandates that \textit{consent must be given through a clear affirmative action} and \textit{consent must be opt-in consent, ..., there is no such thing as `opt-out consent'}~\cite{ico_valid_consent}.
Such a misleading implementation causes users to receive unwanted communications.


In scenario 2, Alice visits \textit{\url{www.giosg.com}} and requests a demo on this website, which leads to the form as shown in Figure~\ref{fig:scenarios_1}. 
The form asks for Alice's personal information, including name, email, phone number, etc. 
The checkbox at the bottom bundles two distinct purposes: \textit{policy agreement} and \textit{consent for marketing communications}. 
This violates the GDPR's requirement for granular consent~\cite{GDPR_Recital_32}, as Alice cannot agree to the privacy policy without simultaneously consenting to marketing communications.

Additionally, both forms in Figure~\ref{fig:scenarios_0} and Figure~\ref{fig:scenarios_1} lack information about consent withdrawal, violating the requirement that \textit{``Withdrawing consent must be as easy as giving it.''}.
In both cases, it is hard for users to withdraw the consent granted to the website, undermining the ease and accessibility of privacy consent policies mandated by the GDPR.

In scenario~3, Alice visits \textit{\url{6sense.com}} and books a demo (Figure~\ref{fig:scenarios_2}. 
The text at the bottom indicates that the company \textit{may occasionally send marketing communications...}, resulting such messages as an inevitable consequence of form submission. 
This design prevents Alice from making a genuine choice, violating the GDPR requirement Recital 42~\cite{GDPR_Recital_42}. 
Without an opt-in mechanism, Alice is forced to accept marketing communications as a condition for accessing the service, which compromises the voluntariness of her consent.

In scenario 4, Alice is browsing \textit{\url{www.adview.com}} and decides to subscribe. 
She finds a subscription form in the website footer (Figure~\ref{fig:scenarios_3}). 
However, the form doesn't state what content she will receive or who will send it (the website or third parties).
As a result, Alice has no way of knowing what she is consenting to. 
This violates GDPR's consent requirements under Article 7~\cite{GDPR_Art7}, which mandates that consent must be \textit{informed} and \textit{specific}. 
Alice cannot give valid consent without knowing the processing purposes and data controller, as she lacks essential information.

\begin{figure}[t!]
    \centering
    \includegraphics[width=.3\textwidth]{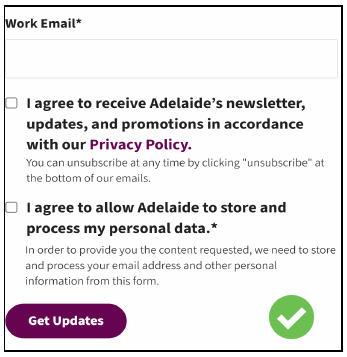}
    \caption{An example of a web form meets all GDPR consent requirements.}
    \label{fig:meet_all}
\end{figure}

In contrast, Figure~\ref{fig:meet_all} shows a compliant consent web form. 
The form uses separate, unticked checkboxes to distinguish optional marketing from necessary services. 
Users are clearly informed about data usage and how to withdraw consent at any time. This design meets all GDPR consent requirements~\cite{GDPR_Art7} regarding freely given, specific, informed, and unambiguous.

\section{Methodology}
\label{sec:method}

In this section, we present the design of \tool that aims to provide an automated reasoning\footnote{\newtext{Note that \textit{automated reasoning} refers to the use of computational logic to provide guarantees about what a system will or will not do, typically through formal inference and proof, which does not imply that all steps of a framework are fully automated~\cite{aws_automated_reasoning}.}}~\cite{automated_reasoning} technique that validates the GDPR compliance on websites.
Specifically, \tool targets to check if the web forms comply with the consent requirements in GDPR. 
We seek a solution that (1) precisely extracts consent mechanisms from a website and (2) automatically validates and localizes GDPR violations.

\begin{figure*}[h]
    \centering
    \includegraphics[width=\textwidth]{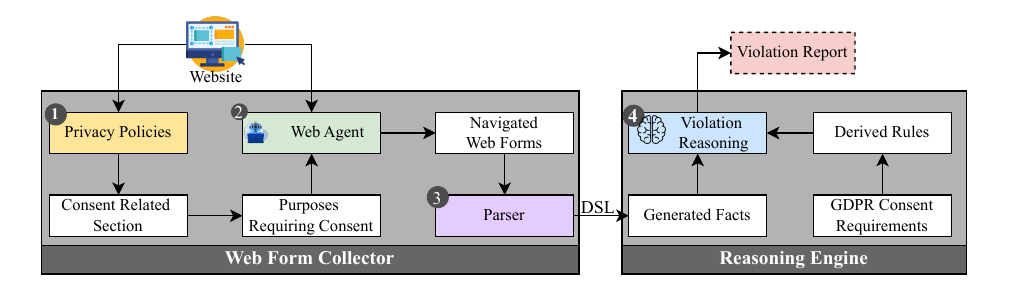}
    \caption{The overall workflow of \tool, including \ding{182} extracting privacy policies on a website, \ding{183} invoking a web agent to collect consent forms guided by the privacy policy, \ding{184} parse the forms into logical facts using the DSL, and \ding{185} validating the privacy violations using the facts against GDPR consent requirements expressed in Datalog rules.}
    \label{fig:system_design}
\end{figure*}

\subject{Overview.} 
Figure~\ref{fig:system_design} presents the overall workflow of \tool.
Given a website as input, \tool
automatically verifies whether any GDPR consent violations exist and generates corresponding violation report.
\tool starts by extracting web forms that require users' consent, which is handled by the \textit{Web Form Collector} module.
To identify the consent-related web forms, it is necessary to track whether and when consent is required in the privacy policy.
Therefore, \tool operates through the following steps:
\ding{182} \textit{Privacy Policy Analysis (PPA):} the \textit{Web Form Collector} collects the privacy policy from the target website and extracts the specific purposes that require user's consent. 
\ding{183} \textit{Web Form Extraction (WFE):} Guided by the identified purposes, a web agent precisely locates and extracts consent-related web forms.
\ding{184} \textit{DSL Encoding:} On top of the obtained web form, \tool parses it to a formal representation using a Domain Specific Language (DSL) to uniformly denote the diverse web forms.
This DSL representation is then fed into the \textit{Reasoning Engine}, which performs:
\ding{185} \textit{Violation Reasoning:} The parser's output is transformed into logical facts, which are subsequently evaluated against GDPR consent requirements (derived as Datalog rules) that have been translated to automatically detect violations. 
The rules explanation with examples was discussed with a legal expert to ensure accuracy.  
Finally, \tool generates a comprehensive violation report, including where the violation occurs and which consent item is violated, if any violations are detected.

\subsection{Privacy Policy Analysis}
As discussed previously, before extracting web forms, we need to determine when consent is required by analyzing privacy policies. However, this presents key challenges due to the unstructured text in the privacy policy and the complexity of identifying consent-related content. Privacy policies are often long documents (7,774 words on average~\cite{nordvpn2023privacy}), and websites may embed or link to additional policy documents across different web pages. 
To address these challenges, we design a multi-step approach that systematically identifies and analyzes consent-related content from privacy policies. PPA consists of three key components: (1) privacy policies collection, (2) consent-related content identification, and (3) purpose extraction to determine specific data processing activities requiring consent.



\subject{Privacy Policies Collection.} To collect privacy policies on a website, \tool first analyzes hyperlinks on the homepage. 
This step is grounded that
\textit{Every organization that maintains a website should publish their privacy notice there, under the title ``Privacy Policy'', and it should be accessible via a direct link from every webpage.}~\cite{gdprprivacynotice} from GDPR.
\newtext{Initially, we only considered links labeled ``privacy policy'' based on this requirement, but observed that websites may also use alternative link labels such as ``privacy'', ``legal'', or ``terms''. We therefore expanded the keyword set to the four to ensure coverage.} However, this broader set may introduce irrelevant pages, \tool applies a two-stage filtering process: it recursively follows identified links to exhaust all potentially relevant pages, and then validates each discovered document by checking its content for privacy-specific terminology such as ``privacy policy'', ``privacy statement'', and ``privacy notice''.

\subject{Consent-related Content Identification.} 
Privacy policies are typically lengthy documents that would be computationally expensive to process entirely with LLMs. 
Therefore, we use Retrieval Augmented Generation (RAG)~\cite{lewis2020retrieval} to extract consent-related content. 
Specifically, we construct a knowledge base by embedding the collected privacy policies and indexing them into a vector database. 
We also use a diverse set of prompts to cover more consent dimensions, e.g.,
\textit{``what user actions are required to give consent''}, \textit{``how users can withdraw consent''}, \textit{``whether consent is cited as a legal basis''}, and \textit{``for what purposes consent is requested''}.


\subject{Purpose Extraction.}~\label{subject:pe} 
Since directly extract purposes from lengthy privacy documents often yields noisy results, we employ a two-stage classification approach with the retrieved content: 
(1) \tool first filters the contents related to consent and excludes the others.
(2) \tool further extracts the specific consent purposes using LLM with in-context learning.
These extracted purposes serve as input to guide automated form extraction in the next stage.
We provide the complete set of prompts in Appendix~\ref{appendix:prompt} Figure~\ref{fig:purpose_extraction_prompt}.


\subsection{Web Form Extraction}
\label{subsec:wfe}
With identified purposes from PPA, WFE aims to extract consent forms from the website. 
However, consent forms may be located on any page without clear indicators, and websites often exhibit diverse structural layouts, embedding forms within iframes, hiding them behind modal dialogs, or requiring user interactions to reveal them. 
For example, the web form in Figure~\ref{fig:scenarios_0} appears on a subscription page, whereas the one in Figure~\ref{fig:scenarios_1} is shown only after a user requests a demo, and the one in Figure~\ref{fig:scenarios_3} is shown in the footer of the website.
The fundamental challenge here is that there are no fixed or unified patterns for web pages to contain forms.
Due to the non-deterministic nature, traditional approaches like using web crawlers to collect data and identify targets with special patterns are not suitable.
Therefore, \tool must adopt adaptive exploration strategies to systematically locate consent forms across diverse websites.

Moreover, extracting each form together with its contextual information is non-trivial, as the relevant semantics may reside outside the \texttt{<form>} tag and be dispersed across multiple DOM elements beyond its boundaries.

To address these challenges, we design WFE with two key components: 
(1) an LLM-based web agent that performs goal-driven (where the identified purposes serve as goals) navigation to automatically explore websites and locate consent forms based on the identified purposes (Section~\ref{subsec:nav}); 
and (2) a multimodal form extraction method that combines HTML parsing with visual analysis to capture both the structural elements of the form and the contextual page information (Section~\ref{subsec:cae}).

\subsubsection{Web Navigation} 
\label{subsec:nav}


To locate consent forms, we build an LLM-based web agent for interactive website navigation based on WebArena~\cite{zhou2023webarena}. The agent operates on the accessibility tree to reason about page content and actions. However, real-world websites introduce challenges beyond WebArena's controlled environment,
 elements are often dynamically rendered, embedded within iframes, or obstructed by overlays and modal dialogs.


To overcome this, we augment the agent with a rule-based prioritized interaction strategy. Upon page load, the agent proactively searches and clicks elements labeled ``Reject All,'' ``Deny All,'' ''Accept All,'' or attempts other dismissal buttons when such global options are unavailable, to dismiss cookie banners and overlays. For non-standard modal dialogs where "close" appears as plain text near unlabeled clickable elements (like \texttt{<span>}), the agent identifies these textual cues and targets the associated visual element, enabling robust dismissal even when conventional labels are missing.


The agent navigates websites using goal-driven instructions derived from identified consent purposes, i.e., ``locate consent form with the purpose of [A]'', where [A] is the purposes identified in Section~\ref{subject:pe}. 
At each step, it selects the next action based on the accessibility tree and the final goal, where the action includes `click', `type', `scroll', `go back', `go forward', `goto(url)', till it finds the web page that include the consent form with purpose [A] or when a stopping threshold is met. During each step, all different modality representations of the web page are saved (including accessibility tree, HTML, and screenshot of the page). The prompt used for web navigation is shown in Figure~\ref{fig:webnav_prompt} in Appendix~\ref{appendix:figure}.

\subsubsection{Consent Form Extraction}
\label{subsec:cae}

An intuitive approach to extract web forms is to locate the HTML \texttt{<form>} tag.
However, this approach is not always practical since web forms can be implemented beyond using the \texttt{<form>} tags~\cite{cui2025webform}. 
For example, the newsletter subscription in \textit{\url{www.shein.co.uk}} directly asks users to input using an \texttt{<input>} tag.
Moreover, analyzing privacy violations would require extra contextual information other than the form itself.
Such information like form descriptions may be located outside the form but rooted in nearby DOM nodes.
For example, the form in \textit{\url{https://adpearance.com/analysis/}} is embedded in an \texttt{iframe} but the descriptions are outside.

To overcome these limitations, we propose a novel approach that uses the Vision-Language Model (VLM) to extract the consent form. 
Our method begins by segmenting the screenshots that contain candidate web forms into smaller figures, which are captured by the web agent as discussed in the previous step (Section~\ref{subsec:nav}).
The reason for having this step is that VLMs would struggle to precisely extract information from large figures due to the resolution limitation and reduced spatial focus~\cite{cai2024spatialbot}.
With the segmented subfigures, we then instruct VLMs to extract form elements (if any), such as input fields and their related text, like labels, descriptions, and button text.
We use these extracted elements as indicators for fuzzy matching.
Finally, we employ a form-matching algorithm in Algorithm~\ref{alg:element_matcher} that precisely extracts the web form from the HTML.
More details can be found in Appendix~\ref{appendix:wfe}.

\begin{figure}[h]
    \centering
    \includegraphics[width=0.48\textwidth]{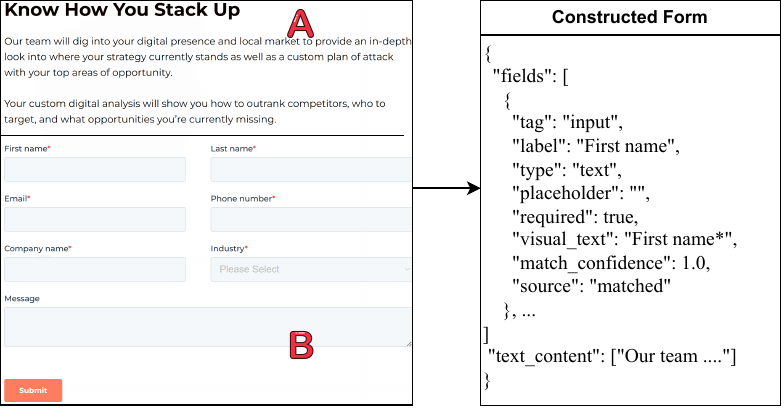}
    \caption{Example of a contextualized form representation. The matched input fields (B) are rendered inside an iframe, while the explanatory text (A) lies outside iframe boundary.}
    \label{fig:contextualized_form_example}
\end{figure}

Taking the example in Figure~\ref{fig:contextualized_form_example}, where Part B is within an iframe, whereas Part A is outside. \tool iterates through HTML elements like the "First Name" input field, computing matching scores with visual elements using type and text matching. The algorithm identifies the best match between the HTML input and visual "First name*" label with high confidence (1.0). Unmatched visual components like explanatory text (A) are extracted as static context, while matched form fields (B) retain their structural attributes including tag type, label, and required status. The final merge creates a unified representation preserving both iframe form structure and surrounding semantic context.

\subsection{Domain Specific Language}
\label{sec:dsl}

In this section, we introduce a Domain Specific Language (DSL) to denote the extracted forms uniformly, which can ease the later compliance verification against GDPR consent regulations. 
Since consent forms lack standardized implementation patterns, the implementation often varies a lot. 
For example, Figure~\ref{fig:scenarios_1} uses elements such as checkbox and dropdown to collect the user's personal information, while Figure~\ref{fig:scenarios_2} only uses textbox. 

Although the forms differ a lot, they always share a similar intention as they are designed to collect users' information and acquire consent. 
This variation brings challenges for systematic compliance analysis due to the diverse implementations of consent forms.
Furthermore, GDPR regulations were written in natural language, which are inherently machine incompatible, making detecting the violations more difficult. 
To address these challenges, we introduce a unified formal representation using a DSL.
In the following sections, we (1) introduce the context-free grammar of the DSL and demonstrate how it formally represents a consent web form, and (2) rigorously formalize the GDPR regulations into Datalog rules in collaboration with domain experts.
\subsubsection{Context-Free Grammar} 
\label{sec:dsl_form}
\begin{figure}[t]
    \small
    \begin{align*}
        \textit{WebForm} &::=\;\; \textit{Content} \\
        \textit{Content} &::=\;\; \textit{Content}+\textit{Content} \mid \textit{Item}  \\
        \textit{Item} &::=\;\; \textit{uid} + \textit{Type} + [ \textit{Metadata} ] \\
        \textit{Type} &::=\;\; \textit{textbox} \mid \textit{button} \mid \textit{checkbox} 
        \mid \textit{combobox} \mid \textit{toggle} \mid \cdots \\
        \textit{Metadata} &::=\;\; \textit{Metadata} + \textit{Metadata} \mid \textit{Action} \mid \textit{Purpose} \\
        &\;\; \quad\!\mid \textit{Controller} \mid \textit{Withdrawal} \mid \textit{RequestText} \\
        \textit{RequestText} &::=\;\; \textit{text} + [ \textit{Polarity} ]\\
        \textit{Polarity} &::=\;\; \textit{affirmative} \mid \textit{negative}
    \end{align*}
    \caption{DSL for consent Web Form descriptions.}
    \label{fig:dsl_syntax}
\end{figure}


Figure~\ref{fig:dsl_syntax} illustrates the context-free grammar of our DSL for modeling the consent web form. 
Specifically, a \texttt{WebForm} consists of one or more \texttt{Items}. 
Each \texttt{Item} has a unique identifier (\texttt{uid}) and a \texttt{Type} (e.g., \texttt{textbox}, \texttt{button}, or \texttt{checkbox}), where a valid \texttt{Type} aligns with role labels in the accessibility tree\footnote{\url{https://developer.mozilla.org/en-US/docs/Glossary/Accessibility_tree}} (e.g., ARIA roles)~\cite{w3c_accessibility_tree}.  
Optionally, an \texttt{Item} may include one or more \texttt{Metadata} fields that capture consent-relevant semantics, including \texttt{Action}, \texttt{Purpose}, \texttt{Controller}, \texttt{Withdrawal}, and \texttt{RequestText}.
All fields of \texttt{Metadata} are \texttt{String}s. 
In particular, \texttt{RequestText} records the textual phrase used to ask user consent, which can be phrased either \emph{affirmatively} (e.g., “I agree to...”) or \emph{negatively} (e.g., ``Check this box if you do not want...'').

For example, we present a part of the representation of the form in Figure~\ref{fig:scenarios_0} using the DSL.
\[\small{
    \begin{aligned}
        WebForm &::= Content + Content + ...\\
        Content &::= Item_1 + Item_2 + Item_3 + ...\\
        Item1 &::= uid_0 + staticText + Metadata_1\\
        Item2 &::= uid_1 + textbox + Metadata_2\\
        Item3 &::= uid_2 + checkbox + Metadata_3\\
        &\cdots\cdots \\ 
    \end{aligned}
}
\]
We use \texttt{Item1} to denote the description of Figure~\ref{fig:scenarios_0} (``\emph{Become a founding member...}'') and \texttt{Item2} for the first input textbox (``\emph{First Name}'').
For the checkbox next to the sentence: ``\emph{Hearst UK... Please tick this box if you'd rather \textbf{not} receive these emails}.'', we use $\texttt{Item3}$ with $\texttt{Type = checkbox}$ and associate it with \texttt{$Metadata_3$}.
We expand \texttt{$Metadata_3$} as below.
\[\small{
    \begin{aligned}
        Metadata_3 &::= Action + Controller + Purpose + RequestText \\
        Action &::= \text{``tick this box''} \\
        Controller &::= \text{``Hearst UK''} \\
        Purpose &::= \text{``email you about our products ...''}\\
        RequestText &::= \text{``Please tick this box if ...''} + [negative] \\
    \end{aligned}
}
\]
In this case, the phrasing in \texttt{RequestText} is negative, potentially obscuring user intent and raising compliance concerns under regulations requiring explicit consent.
This example illustrates how our DSL captures both the web form structure and the semantics necessary for automated consent reasoning.

\subject{Facts Generation.} To enable automated reasoning with Datalog, \tool transforms the DSL representation of a form into a set of Datalog facts, serving as the knowledge base for violation detection. 
\tool first extracts a set of \textit{Base Facts} from the web form in DSL representation. 
The base facts take the form of $\textcolor{blue}{\texttt{pred}}(ID, X_1, X_2, \cdots, X_n)$,
where $\textcolor{blue}{\texttt{pred}}$ is the predicate name representing a specific form component, $ID$ is the unique identifier of the \texttt{Item}, and $X_1, X_2, \cdots, X_n$ denote the relevant information associated with the predicate (if any).
For example, the textbox in Figure~\ref{fig:scenarios_0} is encoded as $\textcolor{blue}{\texttt{item}}(\textit{uid}_0, \text{textbox}, \text{First Name})$, and the checkbox status is encoded as $\textcolor{blue}{\texttt{selected}}(\textit{uid}_2)$. 


In addition to the base facts, \tool derives \textit{Semantic Predicates} by using LLMs to process the static text fields, which are written in natural languages.
For instance, the \texttt{RequestText} ($R$) ``\emph{Hearst UK... Please tick this box if you'd rather not receive these emails}'' in Figure~\ref{fig:scenarios_0} has an action ($A$) ``\emph{tick this box}''.
We thus derive a new \textit{Semantic Predicates} $\textcolor{RedViolet}{\textit{Action}}(R, A)$.
We present the used prompts for semantic predicates and a concrete example in Appendix~\ref{appendix:prompt}.



\begin{table}[t]
    \caption{Semantic Predicates.}
    \centering
    \begin{tabular}{p{0.14\textwidth}p{0.30\textwidth}}
    \toprule
    \textbf{Semantic Predicate} & \centering\arraybackslash \textbf{Intuitive Meaning} \\
    \midrule
    \textcolor{RedViolet}{\textit{Action}}(R, A) & Consent request text $R$ has action $A$ \\
    \midrule
    \textcolor{RedViolet}{\textit{Purpose}}(T, P) & Text $T$ has purpose $P$ \\
    \midrule
    \textcolor{RedViolet}{\textit{ActionMapping}}(E, A) & Action $A$ is mapped that should operate on element $E$ \\
    \midrule
    \textcolor{RedViolet}{\textit{Category}}(T, C, P) & Text $T$ has purpose $P$, which has been classified into category$C$ \\
    \midrule
    \textcolor{RedViolet}{\textit{Controller}}(C) & Controller $C$ is specified on the form \\
    \midrule
    \textcolor{RedViolet}{\textit{Withdraw}}(T, M) & The text $T$ has withdrawn method $M$ \\
    \midrule
    \textcolor{RedViolet}{\textit{Polarity}}(R) & The polarity of consent request $R$ is affirmative \\
    \bottomrule
    \end{tabular}
    \label{tab:semantic_predicates}
\end{table}

\subsubsection{Inference Rules}
\label{sec:dsl_rules}
To reason about the compliance of web consent forms, we define a set of semantic properties derived from GDPR consent requirements, following the official guidance from the UK Information Commissioner's Office~\cite{GDPR_Consent_Requirements, ico_valid_consent}. 
To bridge the gap between GDPR requirements and concrete web form implementation, we formalize each compliance property in GDPR using our DSL and semantic predicates. 
As detailed in Table~\ref{tab:rule}, our DSL-to-GDPR correspondence is grounded in specific regulatory provisions: each property directly maps to GDPR articles, recitals, or official ICO guidance with original expression. 
The formalized properties in Table~\ref{tab:rule} cover \newtext{common} consent requirements defined in GDPR consent~\cite{gdprconsent}, including \textit{freely given} ($P_1$-$P_4$), \textit{specific and informed} ($P_5$-$P_7$), and \textit{unambiguous}($P_8-P_9$).
The mapping from natural language to formal representation in DSL was manually aligned with these provisions in consultation with a legal expert. \newtext{Importantly, this was a one-time effort and the resulting correspondence can be reused across all analyses.}
To reduce syntactic clutter, we use wildcard (\texttt{\_}) for variables that appear only once in a rule. In the following, we briefly describe the motivation behind each property and highlight key aspects of its formal encoding.
\begin{table*}
      \centering
      \caption{Nine consent properties with regulatory descriptions, semantic interpretations, and formal rule specifications.
      }
      \small
      \begin{tblr}{
          colspec = {X[1]X[1]X[1]},
          cell{2}{1} = {r=2}{} , 
          cell{4}{1} = {r=1}{} , 
          cell{5}{1} = {r=1}{} ,
          cell{6}{1} = {r=1}{} , 
          cell{7}{1} = {r=2}{} , 
          cell{9}{1} = {r=1}{} , 
          cell{10}{1} = {r=1}{} , 
          vlines,
          hline{1,11} = {-}{0.08em},  
          hline{2,4,5,6,7,8,9,10} = {-}{}, 
          hline{3} = {2-3}{}, 
          hline{7} = {2-3}{}, 
      }
      GDPR Regulation & Property & Formalized Rules \\
      \textit{Consent means giving people genuine choice and control over how you use their data. If the individual has no real choice, consent is not freely given and it will be invalid.} (\cite{ico_valid_consent}, GDPR Recital 42~\cite{GDPR_Recital_42}) 
      & $\mathcal{P}_1:$ The form text explains the action to provide consent (e.g., click `subscribe') for a specific purpose, and there is an actionable web element corresponding to the action mentioned in the text. 
      & \pone \\
      & $\mathcal{P}_2:$ Selectable elements (e.g., options, checkboxes, binary choices) are used to represent consent agreements. 
      & \ptwo \\
      \textit{There must be a positive action that clearly shows agreement to a specific purpose. Implied methods are only acceptable when obvious and necessary. }(\cite{ico_valid_consent}) 
      & $\mathcal{P}_3:$ If there is only one specific purpose in the web form, then submitting the form shows consent. 
      & \pthree \\
      \textit{Consent should be unbundled from other terms and conditions (including giving separate granular consent options for different types of processing) wherever possible.} (GDPR Article 7(4)~\cite{GDPR_Consent_Requirements}, Recital 43~\cite{GDPR_Recital_43}) 
      & $\mathcal{P}_4:$ Operating on one element should only serve one purpose for consent. 
      & \pfour \\
      \textit{Consent should not be regarded as freely given if the data subject is unable to refuse or withdraw consent without detriment. Informed consent means the data subject knows that they can withdraw their consent at any time. }(\cite{GDPR_Recital_42}, \cite{gdprconsent}) 
      & $\mathcal{P}_5:$ The consent could be withdrawn by the data subject should be illustrated when giving consent. 
      & \pfive \\
      \textit{For consent to be informed, the data subject should be aware at least of the identity of the controller and the purposes of the processing. }(GDPR Recital 42~\cite{GDPR_Recital_42}) 
      & $\mathcal{P}_6:$ On the form, the data controller should be specified. 
      & \psix \\
      & $\mathcal{P}_7:$ On the form, the data processing purpose should be specified. 
      & \pseven \\
      \textit{Consent should be given by a clear affirmative act ... Silence, pre-ticked boxes or inactivity should not constitute consent.} (GDPR Recital 32~\cite{GDPR_Recital_32}) 
      & $\mathcal{P}_8:$ Pre-selected selectable elements showing consent is not allowed. 
      & \peight \\
      \textit{All consent must be opt-in consent - a positive action or indication. Failure to opt out is not consent.} (\cite{ico_valid_consent}) 
      & $\mathcal{P}_9:$ The polarity of consent request should be affirmative instead of negative. 
      & \pnine \\     
      \end{tblr}
      \label{tab:rule}
\end{table*}

$\mathcal{P}_1$-$\mathcal{P}_4$ enforce that consent is freely given, as required by GDPR. 
Specifically, $\mathcal{P}_1$ checks whether the consent request text includes an explicit user action (e.g., ``click the \textit{submit} button'' or ``tick the box'') and whether this action is linked to a concrete web form component. 
This is captured using $\textcolor{RedViolet}{\textit{Action}}(R, A)$ and $\textcolor{RedViolet}{\textit{ActionMapping}}(E, A)$. 
$\mathcal{P}_2$ targets consent designs where a selectable element (e.g., checkbox) is present but the surrounding text lacks any action prompt. 
We detect this using $\textcolor{RedViolet}{\textit{Item}}(E)$ and $\textcolor{RedViolet}{\textit{IsInSelectType}}(E)$.
$\mathcal{P}_3$ only allows implicit consent when a form if and only if one clear purpose (e.g., sending one-time notification) is demonstrated.
$\mathcal{P}_4$ verifies that each selectable element controls only one purpose.
Violations are detected by examining whether multiple $\textcolor{RedViolet}{\textit{Category}}(T, C, P)$ entries with distinct $C$ are associated with the same web form component. 
For example, a single checkbox with the label \textit{``I agree to receive emails and share my data with third parties''} combines purposes from both ``communication'' and ``third-party sharing'' categories, violating the granularity requirement.

$\mathcal{P}_5$-$\mathcal{P}_7$ address informed and specific consent. 
$\mathcal{P}_5$ verifies whether users are informed of their right to withdraw consent (e.g., ``\textit{You can unsubscribe at any time}'') using $\textcolor{RedViolet}{\textit{Withdraw}}(T, M)$, which captures withdrawal-related phrases and their corresponding methods.  
$\mathcal{P}_6$ verifies whether the data controller is disclosed (e.g., ``\textit{XYZ Ltd. processes your data}'') via $\textcolor{RedViolet}{\textit{Controller}}(C)$.  
$\mathcal{P}_7$ checks whether a data processing purpose is explicitly stated using $\textcolor{RedViolet}{\textit{Purpose}}(T, P)$.  

$\mathcal{P}_8$-$\mathcal{P}_9$ ensure unambiguous consent. 
$\mathcal{P}_8$ flags any consent elements that are pre-selected by default (e.g., checkboxes that are already ticked) using $\textcolor{RedViolet}{\textit{Selected}}(E)$. 
$\mathcal{P}_9$ checks whether the request phrasing is in negative polarity; negative forms like ``\textit{Uncheck this box if you don't want...}'' are flagged when $\textcolor{RedViolet}{\textit{Polarity}}(R)$ does not. 

    

\subsection{Violation Reasoning}
\label{sec:violation_reasoning}
\begin{table*}[t]
    \caption{Violation Patterns.}
    \centering
    \begin{tabular}{p{0.13\textwidth}p{0.17\textwidth}p{0.62\textwidth}}
    \toprule
    \textbf{GDPR Principle} & \textbf{Violation Type} & \centering\arraybackslash \textbf{Violation Patterns} \\
    \midrule
    \multirow{2}{*}{Freely Given} & \textit{Genuine Choice} &  $\textcolor{RedViolet}{\textit{Purpose}}(\_, P), \neg(\textcolor{blue}{\mathcal{P}_1}(\_, P, \_); \textcolor{blue}{\mathcal{P}_2}(\_, P, \_); \textcolor{blue}{\mathcal{P}_3}(\_, P, \_))$. \\
    & \textit{Separate Consent} & $(\textcolor{blue}{\mathcal{P}_1}(R, P, E) ; \textcolor{blue}{\mathcal{P}_2}(R, P, E)),
    \textcolor{RedViolet}{\texttt{Category}}(\_, C, P),
    $\\
    & &$ count:\{ C : (\textcolor{blue}{\mathcal{P}_1}(\_, P, E) ; \textcolor{blue}{\mathcal{P}_2}(\_, P, E)),
    \textcolor{RedViolet}{\texttt{Category}}(\_, C, P), \textcolor{RedViolet}{\texttt{Item}}(E, \_, \_) \} > 1$.  \\
    \midrule
    \multirow{3}{*}{Specific and Informed} & \textit{Withdrawal Informed} & $\neg \textcolor{blue}{\mathcal{P}_5}(True)$ \\
    & \textit{Data Controller Specified} & $\neg \textcolor{blue}{\mathcal{P}_6}(C)$ \\
    & \textit{Purpose Specified} & $\neg \textcolor{blue}{\mathcal{P}_7}(P)$ \\
    \midrule
    \multirow{2}{*}{Unambiguous} & \textit{Consent Preselected} & $\textcolor{blue}{\mathcal{P}_2}(\_, \_, E), \textcolor{RedViolet}{\texttt{Selected}}(E)$ \\
    & \textit{Opt-out Consent} & $(\textcolor{blue}{\mathcal{P}_1}(R, \_, E); \textcolor{blue}{\mathcal{P}_2}(R, \_, E)), \neg \textcolor{RedViolet}{\texttt{Polarity}}(R), \textcolor{RedViolet}{\texttt{Item}}(E, \_, \_)$ \\
    \bottomrule
    \end{tabular}
    \label{tab:violation_patterns}
\end{table*}
With Datalog facts extracted from the forms in DSL representation and GDPR requirements formalized as logical predicates (in Section~\ref{sec:dsl}), \tool performs violation reasoning through pattern matching. 
We define the violation patterns based on the GDPR compliance rules in Table~\ref{tab:rule}.

\subject{Defining Violation Patterns.} 
Since GDPR compliance allows multiple valid implementation paths for the same legal requirement, we thus define violation patterns using our DSL based on the compliance rules $\mathcal{P}_i$ in Table~\ref{tab:rule} for violation detection. 
For example, rules $\mathcal{P}_1$, $\mathcal{P}_2$, $\mathcal{P}_3$ represent different web form patterns through which consent can be established (e.g., through explicit action, selectable elements, or implied submission under a strict constraint). 
Since these rules are semantically interchangeable, we treat them as alternative satisfiers, i.e., a violation is reported only if none of these rules are satisfied, which violates \textit{Genuine Choice} in \textit{Freely Given.} 
The complete list of violation patterns is shown in Table~\ref{tab:violation_patterns}. 
For different GDPR consent principles, we define seven types of violations in total.

\begin{figure}[t]
    \centering
    \includegraphics[width=\linewidth]{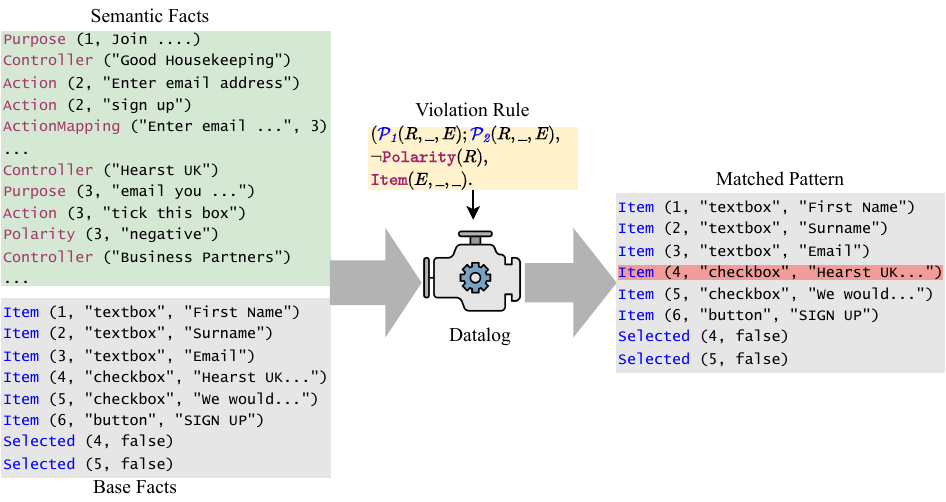} 
    \caption{An end-to-end example of violation detection: given extracted base and context facts, the Datalog engine applies the violation rule (e.g., $\mathcal{P}_8$) to match a pattern where a checkbox uses a negative polarity request text without an affirmative opt-in action.}
    \label{fig:violation_example}
  \end{figure}

\subject{Automated Reasoning and Reporting.} 
\tool applies all defined violation patterns to the facts generated from the web form using a Datalog engine. 
For each matched violation pattern, \tool identifies a violation of a specific GDPR requirement and generates a corresponding structured report.
Each report includes the violation type, the corresponding rule ID (e.g., $\mathcal{P}_7$), the unique identifier of the web element, type (e.g., checkbox \#4), and other relevant textual context (e.g., the \texttt{RequestText} triggering the violation). 
Each violation is localized to the specific web form component involved. 
Figure~\ref{fig:violation_example} illustrates an example of this process, showing how \tool extracts structured facts from a web form, applies a violation pattern (e.g., $\mathcal{P}_8$), and identifies a matched pattern indicating non-compliant behavior.

\section{Experiment}
\label{sec:experiment}
In this section, we describe our implementation and dataset in Section~\ref{sec:exp}. Also, we provide the evaluation of \tool. In the end, we evaluate the end-to-end performance of \tool (Section~\ref{sec: rq1}), measurement analysis on wild websites (Section~\ref{sec: rq2}), and computation overhead in Section~\ref{sec: rq3}.  
\subsection{Experiment Setup}
\label{sec:exp}
\subject{Implementation.} For \textit{Privacy Policy Analysis}, we employ PoliPy~\cite{polipy} to download the corresponding privacy documents and filter out irrelevant links that do not contain the keyword ``privacy'' or ``terms'' in their web page main content. Then, we use LangDetect~\cite{langdetect} to identify the language of each document and discard those that are not in English. 
For RAG, we used Langchain~\cite{langchain} with \texttt{RecursiveCharacterTextSplitter} to process documents, and \texttt{OpenAIEmbeddings} for vectorization. Retrieval is performed using its similarity search with up to 15 chunks per query.  
For \textit{Form Extraction}, we adapt the WebArena framework, integrating with Playwright for automated website navigation and form discovery.
During web navigation, the intent is set using the purpose extracted in PPA, i.e, ``Navigate to the website that contains web form(s) with purpose [A]''. The exploration process terminates when the goal is achieved (i.e., the target form is found) or when stopping criteria are met, including a maximum of 30 total steps or five repeated actions on the same page to prevent infinite loops.
Screenshot segmentation is performed using OpenCV~\cite{opencv}. 
All (V)LLM-based tasks throughout the pipeline employ \texttt{GPT-4o-mini} for consistency. 
For \textit{Violation Reasoning}, we implement the violation rules as formal Datalog queries using Soufflé 2.2 (64-bit Domains)~\cite{souffle}. Each extracted form is first converted into base facts representing structural elements and semantic predicates generated through LLM analysis. The Soufflé engine executes queries against these facts to detect violations.  In total, the implementation of \tool contains 3,110 Lines of Code (LoC) in Python (including the modification on WebArena) and 279 LoC in Soufflé.

\subject{Dataset.}  We used the Tranco dataset\footnote{\url{https://tranco-list.eu/list/KJYKW}} from the list dated 11 Feb, 2025. To ensure broad coverage across diverse website categories, we selected the top 50,000 websites from this list as our initial dataset. We employed \textit{Good-Turing frequency estimation}~\cite{gale1995good} to estimate the number of websites required to represent the majority of categories. We retrieved category labels using the Categorify API~\cite{categorify}, which classifies websites into 70+ distinct categories. 
After excluding websites that were not accessible at the time of analysis, we obtained 41,603 valid URL-category pairs (If a website covers multiple categories, it is counted multiple times). Based on the estimation, we found that analyzing 12,355 websites would suffice to cover 97\% of the observed categories. Therefore, we selected and analyzed the top 12,355 visitable websites from the dataset to ensure representative category coverage in our study. After filter those websites that are not in English or have no privacy policy, 5,823 websites remained for analysis. 
Further, since our analysis focuses on GDPR compliance, and the requirements for consent vary across different countries, we perform the analysis by accessing the websites from the United Kingdom via Bright Data Proxy~\cite{brightdata2021} to ensure the  content retrieved reflects GDPR-relevant implementations.

\subject{Experiment Settings.} We ran \tool on a server Ubuntu 22.04.1 LTS, equipped with 16-core Intel Xeon Gold 5222 CPUs, 125 GiB of RAM, and over 17 TB of total storage (including local and network-mounted volumes).

\subject{Research Questions} To evaluate the effectiveness of \tool, we seek answers to the following questions:
\begin{itemize}
    \item RQ1: What is the end-to-end performance of \tool in detecting GDPR consent violations in consent mechanisms?
    \item RQ2: How prevalent are GDPR consent violations in the wild?
    \item RQ3: What is the computation overhead of \tool?
\end{itemize}

\subsection{End-to-End Performance (RQ1)}
\label{sec: rq1}
To evaluate the ability of \tool to identify consent violations, we compare the results of \tool with manual analysis results. 
We \newtext{randomly sampled}  150 web forms across 150 websites. 
\newtext{For the end-to-end evaluation, each case begins from the website's URL.}
In the following, we report the results of \tool by evaluating these 150 web forms.

We report the true postive (TP), false positive (FP), true negative (TN), and false negative (FN) rates of the analysis in Table~\ref{tab:performance}. We consider the violations as positive data and non-violations as negative data. TP indicates the consent mechanism is labeled by \tool and human annotator to be in violation, FP are labeled by \tool in violation whereas manual analysis determines them to be compliant (non-violation), TN denotes the consent mechanism is labeled by both \tool and human annotator to be compliant, and FN indicates the consent mechanism is labeled by \tool as compliant but manual analysis determines it to be in violation. The manual annotation was conducted by two authors independently. Inter-annotator reliability was assessed using Cohen's Kappa ($K = 0.861$). The disagreements between the annotators were resolved through discussion until consensus was reached. Additionally, one-third of the annotated data was further validated by a legal expert to ensure accuracy.

\begin{table}
    \centering
    \caption{Detailed performance of \tool on 150 web forms for identifying violations. Here, TP=True Positive, TN=True Negative, FP=False Positive, FN=False Negative.}
    \label{tab:performance}
    \begin{tblr}{
      vline{2,6} = {-}{},
      hline{1-2,9} = {-}{0.07em},
      hline{9-10} = {-}{},
    }
    Violation Type            & TP & TN & FP & FN & TPR & TNR \\
    No Genuine Choice            & 1   & 149   &  0  & 0   &  100\%   &  100\%   \\
    Not Separate Consent           &  7  & 141   &  2  &  0  &  100\%   & 98.6\%    \\
    No Withdraw Information       &  98  &  50  &   1 &  1  &  99.0\%    &  98.0\%   \\
    No Controller & 3   &   144 &  3  &  0  &  100\%    &  98.0\%   \\
    Purpose Not Specified         &  1  & 149   &   0 &  0  &  100\%  &  100\%   \\
    Consent Pre-selected       & 12   & 138   & 0   &  1  &  92.3\%  &  100.0\%   \\
    Opt Out Choice       &  1  &   148 &  1  &  0  &   100\%  & 99.3\%    \\
    Total            & 122   &  919  &  4  &  5  &  98.6\%   & 99.1\%     
    \end{tblr}
\end{table}

We observe that \tool achieves an average of 98.6\% TPR and 99.1\% TNR in detecting seven types of violations. In the following, we discuss in detail about the performance in each of these categories.




1) \textit{Evaluation of Genuine Choice}: \tool identified 1 form as violating the requirements of Genuine Choice. \tool achieves 100\% on both TPR and TNR. The main reason for the violation is that the form presents users with a ``take it or leave it'' approach, i.e., users cannot select if they want to receive specific services but must accept all offerings or abandon registration entirely.

2) \textit{Evaluation of Separate Consent}: Out of the 150 web forms from manual analysis, \tool marks 9 forms as violating the separate consent requirements with a TPR of 100\% and a TNR of 98.6\% (141 out of 143). Our tool identified 2 false positives. For example, the form shown in Figure~\ref{fig:sc_fp_case} in Appendix~\ref{appendix:figure} was flagged as violating separate consent by mistake. This is because when extracting semantic predicates using LLM, the tool parsed ``\textit{Subscribe to the RTB House newsletter and get the latest MarTech industry insight, including updates on the cookieless future, ideas for winning ad campaigns, and the great leaps in AI technology}'' and identified multiple distinct purposes categories (newsletter subscription, advertising insights). However, manual analysis revealed these are all descriptions of the same newsletter service, not requiring individual opt-in mechanisms.

3) Evaluation of \textit{Withdrawal Informed}: Out of the 150 web forms from manual analysis, \tool marks 97 forms as violating withdrawal information requirements with a TPR of 99.0\% and a TNR of 98.0\% (50 out of 51). Our tool missed 1 true violation (false negatives). For instance, the Adsquare contact form (Figure~\ref{fig:sc_fp_case} in Appendix~\ref{appendix:figure}) includes withdrawal information in fine print at the bottom of the page. 
However, the text about withdrawl \textit{``You can change your mind at any time ...''} is outside the \texttt{<form>} tag, and the font size is much smaller than main text and visually separated from the main form elements, even VLLM cannot extract the information.
This disconnection in layout and semantics led to a missed violation.

4) Evaluation of \textit{Data Controller Specified}: \tool marks 6 forms as violating data controller specification requirements with a TPR of 100\% and a TNR of 98.0\% (144 out of 147). Our tool identified 3 false positives. These cases primarily arise when controller information is conveyed implicitly through company branding or logos rather than textual statements. Since \tool focuses on extracting textual declarations of the data controller, it may fail to recognize visual cues like logos that are easily understood by human users but not captured by our text-based analysis.

5) \textit{Evaluation of Purpose Specified}: \tool identified 1 form as violating purpose specification requirements. which achieves TNR of 100\% and TPR of 100\%. The main reason for most forms not violating this is that the form often provides clear purposes to attracts users to use their service, e.g., ``for contacting you about our services'' or ``to send newsletter updates''.
 
6) \textit{Evaluation of Consent Preselected}: \tool marks 12 forms as violating consent preselection requirements with a TPR of 92.3\% and a TNR of 100\%. Our tool missed 1 true violation (false negative). The false negative occurred when the form used 
\texttt{<input type="checkbox" class="active">} with CSS classes like \texttt{.active} to control the preselected state through styling, rather than the standard HTML \texttt{checked} attribute. 

7) \textit{Evaluation of Opt Out Choice}: \tool marks 2 form as violating opt-out choice requirements with a TPR of 100\% and a TNR of 99.3\%. The false positive cases mainly stem from unsubscribe forms being misclassified as consent forms. For example, on \textit{\url{massgeneral.org}}, an unsubscribe form was mistaken for a consent form, leading \tool to incorrectly flag the statement “\textit{Unsubscribe my email address from all communications}” with a checkbox as a violation, despite it being the intended withdrawal functionality.



In addition to violation detection performance, to evaluate the accuracy of \tool's pipeline components, we randomly sampled 100 websites to inspect each step of the pipeline. When visiting the websites, we used NordVPN~\cite{nordvpn} to change the location to the United Kingdom. Among the 100 websites, privacy policy extraction achieves 97.91\% recall on English websites (n=48). 52 websites were excluded as they either had no privacy policy or were not in English.
Form navigation success rate is 93.75\%, with failures including 1 CAPTCHA not bypassed and 2 forms not found with claimed purpose. Web form element matching correctly aligns HTML and visual components in 100\% of cases, indicating that all required form information and attributes were successfully extracted.   

\subject{\newtext{Characterization of FPs \& FNs.}} \newtext{As discussed above, parts of \tool rely on LLMs, whose unpredictability can introduce false positives and false negatives. To clarify when such errors may arise, we provide an abstract characterization of the error modes:
(1) \textit{Privacy policy analysis errors.} Cases where consent-related statements or purposes in the privacy policy are either missed or incorrectly extracted. (2) \textit{Form extraction errors.} Form extraction depends on website HTML and VLLM analysis, errors may occur if the VLLM misinterprets the relationship between layout and semantics, leading to relevant text being overlooked or not linked to the form (as discussed in \textit{evaluation of withdrawal informed}). Additionally, some forms are specifically designed for actions such as unsubscribing, but because they share similar purposes with consent forms, they may be incorrectly extracted as consent-related, introducing further errors (as discussed in \textit{evaluation of opt out choice}). (3) \textit{Facts Generation Errors.} In semantic predicate generation, the LLM may over-interpret or hallucinate content beyond what is supported by the original text (as discussed in \textit{evaluation of separate consent}). Nonetheless, all these errors stem from the LLM components and are confined to text processing rather than complex reasoning. Once the form facts are established, the reasoning stage in \tool is fully deterministic, thereby ensuring consistent compliance analysis. As LLMs continue to advance, the frequency of such text-processing errors is expected to diminish.}



\subsection{Measurement Analysis in the Wild (RQ2)} 
\label{sec: rq2}
\begin{figure}
    \centering
    \includegraphics[width=0.48\textwidth]{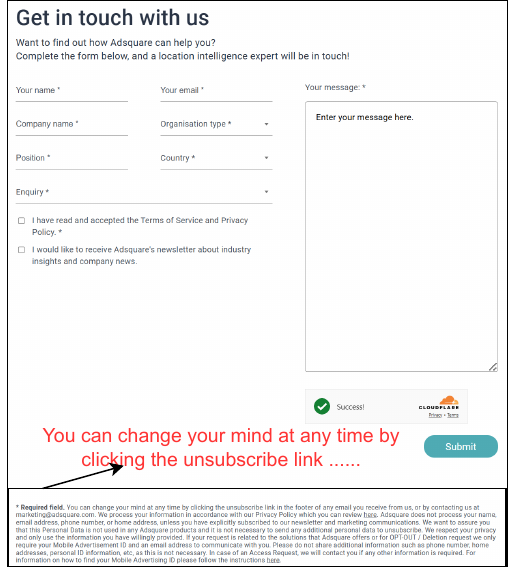}
    \caption{Real world example of a withdrawal violation false positive of \tool. The form is extracted from the website \url{https://adsquare.com/about-us/}}
    \label{fig:w_fp_case}
\end{figure}

\subject{Landscape.} Our study reveals that GDPR consent violations are prevalent among popular websites across diverse categories. Running \tool on 5,823 websites, with 3,599 web forms (using consent) found, \tool reported 3,485 (96.8\%) consent violations across 79 website categories. The violations span various GDPR principles:  3,223 web forms have missing withdrawal information (89.6\%), 286 lack specific purpose disclosure (7.9\%),  49 use pre-selected consent options (1.4\%), 20 employ opt-out consent patterns (0.6\%), 440 do not specify data controllers (12.2\%), and 435 bundle multiple consent purposes inappropriately (12.1\%).
\newtext{
To gain deeper insight into the nature of these violations, we examined the patterns across both compliant and non-compliant forms from Section~\ref{sec: rq1}. Our analysis reveals that violations predominantly result from the complete 
absence of required information. However, even among technically compliant forms, we observed information presentation issues where there is a tendency to de-emphasize important information and make it less apparent to the user. For example, among the 51 compliant cases with withdrawal information available (Table~\ref{tab:performance}), we found 
that 10 out of 51 displayed this information in ways that reduced visibility, using smaller fonts, low-contrast text, or placing it separated from the main consent request (Figure~\ref{fig:w_fp_case}). 
This issue of downplaying important information also applies to other consent elements, such as purposes or data controller details, which further prevent users from exercising their rights in a meaningful way. 
Therefore, ensuring compliance requires not only including the required information but also presenting it clearly and prominently. 
Both UI designers and developers should recognize that subtle presentation details can critically affect the validity of consent.}

To assess consent compliance, we use a two-level approach. Specifically, for each website $i$ with $F_i$ consent forms, we calculate violation rates for seven violation types as $V_{ij}=\frac{\text{forms with violation }j}{F_i}$.  The website violation rate is computed as $\text{Website Rate}_i = \frac{1}{7}\sum_{j=1}^{7}V_{ij}$. At the second level, category violation rates are calculated as $\text{Category Rate}_c = \frac{1}{n_c}\sum_{i=1}^{n_c}\text{Website Rate}_i$ for categories.

Table~\ref{tab:violation_category} illustrates the top 5 website categories with the highest violation rates. \textit{Adult/Pornography} websites exhibit the highest violation rate (36.1\%) violation rate with near-universal non-compliance (98.1\% of websites under category), \textit{Information Security} websites exhibit the best compliance performance across all categories (25.5\% violation rate). A striking pattern emerges across all categories: withdrawal unavailability violations consistently represent the most common compliance failure (90.6\% of websites), regardless of website category.
\newtext{These differences suggest that violation rates are shaped by domain-specific practices: categories such as \textit{Adult/Pornography} may prioritize monetization and provide minimal or no consent controls, whereas \textit{Information Security} websites are more likely to implement complete consent mechanisms including accessible withdrawal mechanisms, reflecting stronger compliance awareness. } 

\begin{table}
    \caption{Top 5 website categories with the highest violation rates.}
    \label{tab:violation_category}
    \centering
    \begin{tblr}{
      hline{1-2,7} = {-}{0.08em},
    }
    Rank & Category & Violation Rate~ & Sites Analyzed \\
    1    &  Adult/Pornography        &    36.1\%             &    53                    \\
    2    &  Cloud Service/Hosting        &    32.0\%              &     91                   \\
    3    &  Health        &   31.7\%              &                33        \\
    4    &  Entertainment        &       29.8\%          &           25             \\
    5    &  Advertising \& Tracking        &     28.9\%            &         45               \\ 
    \end{tblr}
\end{table}

\begin{figure}
    \centering
    \includegraphics[width=.45\textwidth]{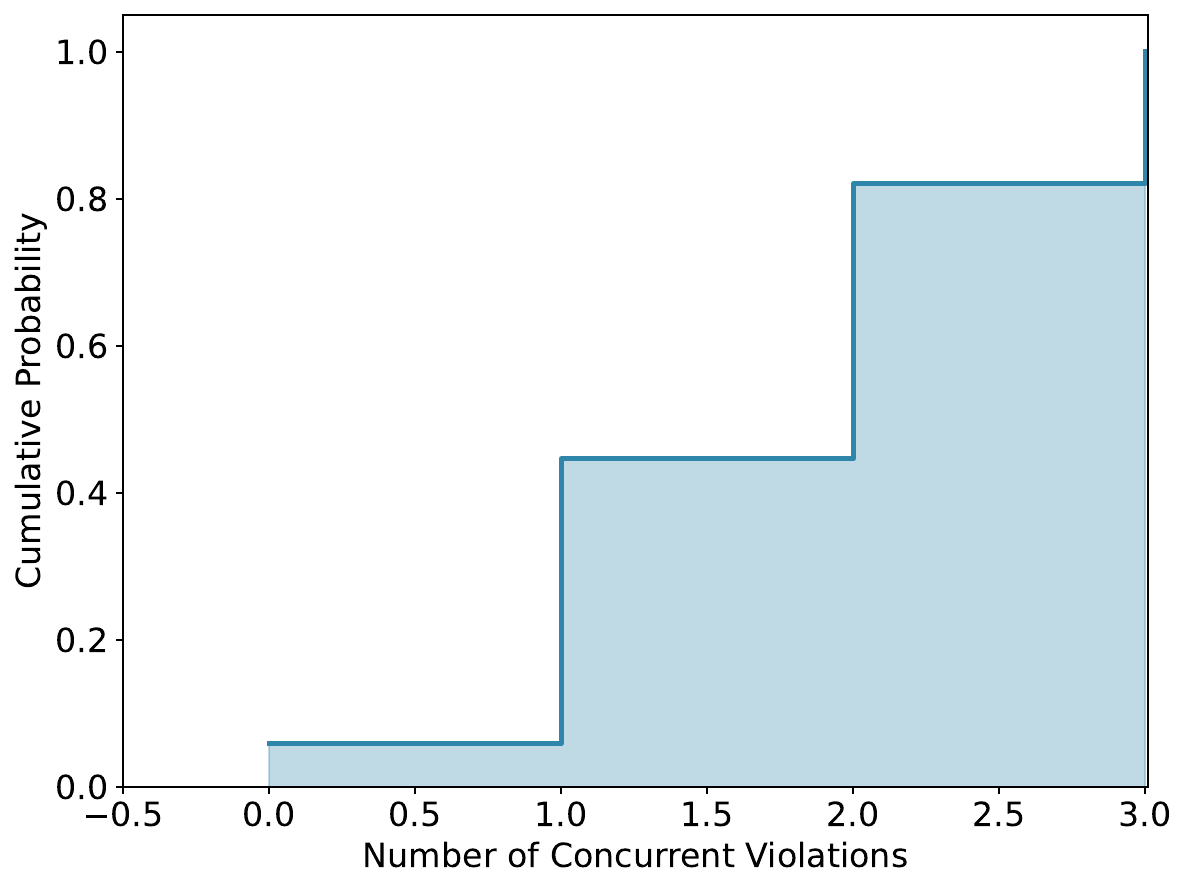}
    \caption{CDF of Violation Co-occurrence.}
    \label{fig:cdf}
\end{figure}

\subject{Degree of Violation Severity.} Some consent mechanisms may exhibit multiple GDPR violations simultaneously, and in case of non-compliance, they may violate consent principles across different categories.  We then assess the severity level of consent violations for websites, specifically analyzing the extent to which multiple violation types co-occur. Figure~\ref{fig:cdf} presents the cumulative distribution function (CDF) of violation co-occurrence patterns across 3,599 consent instances analyzed in our study. The CDF analysis reveals alarming patterns of systematic GDPR non-compliance, with only 5.9\% of instances demonstrating complete compliance and 38.7\% exhibiting single-violation patterns. Critically, 55.3\% of consent mechanisms exhibit multiple 
concurrent violations.

\subject{Personal Data Collection in Violating Forms.} We also analyze the PIIs collected in GDPR consent violations. Examples of such PIIs are email addresses, phone numbers, full names, postal addresses, and financial information. We break down PIIs in the consent violations based on data categories. We show the details in Table~\ref{tab:pii}. Email addresses are collected by the majority of violating forms across all categories: 86.2\% of forms with Freely Given violations, 72.1\% of forms with Specific \& Informed violations, and 82.4\% of forms with Unambiguous violations collect email addresses. Notably, forms with Unambiguous violations show consistently higher PII collection rates across multiple data types—while first and last name collection occurs in only 15-23\% of other violation types, it jumps to 35.2\% each in Unambiguous violations, and phone number collection increases from 14-19\% to 27.5\%.  Beyond the PIIs shown in Table~\ref{tab:pii}, violating forms also collect other sensitive personal data, including postal addresses, financial information (e.g., credit card details, bank account numbers), and demographic information (e.g., age, gender).


\begin{table}
    \centering
    \caption{Distribution of Top 5 PIIs in GDPR consent violations.
    }
    \label{tab:pii}
    \begin{tblr}{
      hline{1-2,7} = {-}{0.08em},
    }
    PII Type & Freely Given & Specific \& Informed & Unambiguous \\
    Email         &   86.2\%           &   72.1\%                 &  82.4\%            \\
    First Name         &     22.6\%          &     15.8\%               &    35.2\%         \\
    
    Last Name         &    21.9\%          &       15.3\%             &   35.2\%    \\
    Phone       &   18.7\%           &      14.4\%                &  27.5\%  \\ 
    Company         &    14.7\%          &     10.9\%               &  13.2\%   \\
    \end{tblr}
\end{table}


\subsection{Analysis Time Distribution (RQ3)}
\label{sec: rq3}
We benchmarked the runtime performance of \tool on the 5,823 websites. The benchmarks ran on the same machines described in Section~\ref{sec:exp}. 
In total, it took around 15 days to finish all tasks, including privacy policy analysis, consent form extraction, DSL encoding, and violation reasoning. 
Note that the analysis was performed in parallel. Additionally, if no consent form was found on a website, the latter two modules were skipped. Therefore, we report runtime performance only for websites where consent forms were extracted.

Overall, the analysis time per website ranged from 6.3 minutes to 37 minutes, with an average of 20.2 minutes.
 Among the three components, \newtext{the \textit{Consent Form Extraction} module is the most time-consuming, taking on average 17.9 minutes per website due to LLM-based long-context processing with the accessibility tree and images for interactive navigation.} \newtext{The \textit{Privacy Policy Analysis} module requires on average 2.3 minutes}.
  In contrast, the VR module demonstrates excellent efficiency with most analyses completing with an average of 0.2 second, as the Datalog-based formal reasoning scales linearly with form complexity. 

\newtext{This analysis time is related to task complexity. The time variance (6.3 to 37 minutes) 
reflects website complexity differences: simpler sites with forms on 
landing pages are completed quickly (less than 10 minutes), while complex sites require 
extended exploration due to multi-level navigation, multiple overlays 
requiring dismissal, or forms embedded behind sequential interactions.  For example, \url{www.007.com} has a compact accessibility tree, requiring 
only one button click to access the consent form, which leads to minimal 
LLM context and completes the total analysis in 6.8 minutes. In contrast, \url{www.shein.com} has multiple consent forms across different pages, 
requiring navigation through several popup dialogs and dynamic content. The large 
accessibility trees and multiple navigation rounds result in extended LLM processing, 
completing in around 30 minutes in total.
}


\section{Discussion}
\label{sec:discussion}

In this section, we summarize a few important discussions related to \tool, including analysis reliability, adaptability, scalability, and limitations.


\subject{Adaptability to other regulations.} \tool's DSL enables extension to other privacy regulations with minimal modification. Adapting to new regulations like CCPA primarily requires updating the violation reasoning rules. For example, CCPA~\cite{ccpa-law} mandates prominent ``Do Not Sell'' disclosures, which could be formalized as new rules.

\subject{Scalability.} \tool easily accommodates new consent mechanisms with minimal effort. 
When incorporating the new ones, only several straightforward changes are needed:
(1) Add new grammar to extend the existing DSL, including new predicates.
For example, for runtime consent mechanisms that dynamically appear based on user's behavior (like location-based consent prompts when users access geo-restricted content), we would add predicates like {\textit{RuntimeConsent}}(trigger\_condition, timing, user\_context) to capture the contextual consent requirements.
(2) Incorporate new Datalog rules, which is just a one-time effort.
For instance, for consent mechanisms that appear during user sessions based on detected activities, we could define \textit{ActivityTriggeredConsent}(activity\_type, consent\_timing, user\_wareness) and rules to verify that runtime consent requests provide the same level of transparency.
This lightweight design allows \tool to quickly adapt to emerging consent practices without affecting existing capabilities.

\subject{UI-Level Analysis Scope.} \tool exclusively analyzes consent mechanisms in the consent form at the user interface level, which provides only a partial view of GDPR compliance. 
The limitation lies in that we cannot verify whether consent choices are actually honored in backend processing, trace data flows to third parties, assess consent record storage, or confirm that withdrawal mechanisms function properly. Consequently, a form may appear compliant at the interface level but still process data without proper consent validation in the backend. 
However, when UI level violations are present, they directly impact users' ability to provide valid consent, making backend compliance irrelevant. 

\subject{Recommendation For Different Stakeholders.} As shown in Section~\ref{sec: rq2}, more than 90\% of the web forms have at least one consent violation, emphasizing the urgent need for action from multiple stakeholders. For \textit{website operators and developers}, it is critical to move away from dark patterns and ensure consent elements are clear, purpose-specific, and accompanied by accessible withdrawal options. For \textit{legal and compliance teams}, relying solely on privacy policy declarations is insufficient; compliance must be enforced at the interface level, where actual user interaction occurs. 

\subject{\newtext{Intended Users and Utility.}} \newtext{\tool is designed to accommodate diverse stakeholders, including website developers, researchers, and regulators, through differentiated modes of operation.  
For website developers, \tool could be used in a lightweight mode, where they directly provide the web forms to the DSL encoder and reasoning engine for development-time compliance checking. In this mode, developers do not need additional features like webform extraction and benefit from a deterministic reasoning process that minimizes false positives, since the analysis is fully reliable once the form facts are specified. For other stakeholders such as researchers and policymakers, \tool's full mode enables end-to-end analysis that starts from a given website URL and encompasses privacy policy analysis to backend compliance verification.}

\subject{\newtext{Limitations.}} \newtext{\tool demonstrates its utility by enabling large-scale detection of GDPR consent UI violations, revealing widespread non-compliance patterns across thousands of websites (Section~\ref{sec: rq2}). However, the results alone cannot explain why developers introduce these issues, since violation statistics cannot reveal whether non-compliance arises from lack of awareness, misinterpretation of regulatory requirements, or competing design priorities. We identify this limitation as an important direction for future work that would require developer interviews to understand the underlying causes and decision-making processes behind these violations.}

\section{Related Work}
\label{sec:related_work}

\subject{Consent Compliance.}  Extensive research has been conducted on consent compliance across various platforms. One major area of focus is cookie consent banners\cite{bollinger2022automating, matte2020cookie, bouhoula2024automated, bielova2024effect, papadogiannakis2021user, bui2022opt, klein2022accept, utz2019informed}, where studies examine the extent to which website cookie banners comply with consent-related regulations. These works investigate banner designs~\cite{bielova2024effect}, evaluate opt-out mechanisms~\cite{bui2022opt}, and assess whether banners genuinely honor users' choices~\cite{matte2020cookie}. Another line of research explores smartphone consent dialogs~\cite{nguyen2021share,nguyen2022freely,koch2023ok, mohamed2024attention} investigated the effectiveness of consent management in mobile applications, demystifying various violations of GDPR's explicit consent. Also, Li et al.~\cite{li2024we} investigated runtime privacy notice practice in mobile apps. While prior studies provide crucial insights into consent for tracking technologies and mobile application permissions,
web forms collect more PIIs, including users' email, age, and phone number. However, their methods cannot be directly applied due to the unpredictable location of web forms and the different representation between web forms and cookie banners. Unlike cookie banners with standardized loading patterns, web forms may surface dynamically through user interactions, rendering existing detection methods inadequate for form-based consent analysis.

\subject{Statement and Practice Compliance.} Privacy compliance checks focus on the consistency between data usage practices and data usage statements (e.g., privacy policy), including data collection, data sharing, and withdrawal practices~\cite{andow2019policylint, andow2020actions, breaux2013formal, slavin2016toward, wang2018guileak, yu2016can, zimmeck2019maps, zimmeck2017automated, bui2023detection, bui2021consistency, du2024withdrawing}. However, these works focus on standalone consent mechanisms (cookie banners, app dialogs) and overlook how consent is solicited within functional web forms.

\subject{Webform.} Web forms are a primary channel for collecting PII online and have been studied from multiple privacy-related perspectives. 
Prior work has examined premature data collection before submission~\cite{senol2022leaky}, large-scale registration and classification of legal properties to detect GDPR violations~\cite{kubicek2024automating}, \newtext{data type extraction~\cite{kafle2024understanding}}, and the derivation of privacy norms from collected information~\cite{cui2025webform}.
\newtext{However, these studies primarily focus on data collection and overlook the semantics of consent in the broader form context. We complement this line of work by modeling consent semantics through a DSL that separately represents form UIs and GDPR rules, enabling automated detection of consent violations via formal reasoning.}

\subject{Privacy policy analysis/GDPR analysis.} 
Early approaches, such as Watanabe et al.~\cite{watanabe_understanding_2015}, relied on keywords and heuristics to identify potential non-compliance, but often struggled with the ambiguity and complexity of legal language. Later work introduced annotated corpora that enabled machine learning techniques: Wilson et al.~\cite{Liu_Wilson_Story_Zimmeck_Sadeh} evaluated basic NLP models, while Harkous et al.~\cite{harkous_polisis:_2018} developed a multi-label CNN classifier. Although these models achieved strong overall performance, they performed poorly on minority classes due to imbalanced training data, a limitation later addressed with synthetic augmentation~\cite{zimmeck2019maps} and with active learning and crowdsourcing in Calpric~\cite{qiu_calpric:_2023}. PolicyLint~\cite{andow2019policylint} and PoliCheck~\cite{andow2020actions} organize information types into a hierarchical ontology to detect conflicts, while Poligraph~\cite{cui2023poligraph} represents privacy policies as knowledge graphs. \newtext{However, while these tools can also extract purposes, they typically rely on explicit sentence structures (e.g., "we collect your name for [purpose]"). In the consent context, however, purposes are often expressed in diverse and implicit forms (e.g., "without consent, we won't [purpose]"), which such approaches fail to capture. By contrast, Cosmic's LLM-based analysis is able to identify these implicit purposes.}

\section{Ethics}

\subject{Impact on Live Systems.} \newtext{Since our study relied on automated web navigation to extract consent forms, we carefully designed the process to minimize risks and avoid harm, guided by the Menlo Report principles~\cite{kenneally2012menlo}. Specifically, we implemented strict rate limiting and bounded exploration depth ($\leq$ 30 steps per site, terminating earlier if repeated actions occurred), which prevented infinite loops and excessive requests. The crawl operated at a low overall rate with limited concurrency, making its traffic volume comparable to or lower than ordinary browsing. We never submitted sensitive or non-public information in forms, and interactions were limited to public-facing consent interfaces. These safeguards ensured that websites were not overwhelmed or misused, in accordance with established ethical guidance for network measurements~\cite{partridge2016ethical}.
}

\subject{Violation Disclosure.} \newtext{Our analysis indicated a high rate of non-compliance across evaluated forms (>90\% of evaluated forms). To assess the feasibility of individual disclosure, we notified 30 website operators of detected violations. So far, only one replied with an acknowledgement, while most have not responded yet. Different from security vulnerabilities, which can often be addressed directly by developers, any response for privacy compliance issues will likely require review by legal counsel instead of just developers, and the process is thus slow. This experience indicates that large-scale notification is impractical. Following the Menlo Report principles of Justice and Respect for Law and Public Interest~\cite{kenneally2012menlo}, we therefore chose not to pursue selective disclosure further. Instead, we report results only in aggregate, focusing on systemic insights that can inform regulators, practitioners, etc. without unfairly burdening individual websites.}

\section{Conclusion}
\label{sec:conclusion}


In this paper, we present a novel automated reasoning technique for validating privacy violations in GDPR's consent requirements and develop it as a software called \tool.
Our insights are threefold:
(1) We introduce a DSL to systematically describe the diverse web forms and denote them using a formal representation; 
(2) We convert the GDPR consent requirements to Datalog rules, bridging the gaps of applying verification technologies to detect privacy violations. 
(3) We encode the web form expressed in the DSL to logical facts and employ Datalog engine for privacy violation detection.
\tool reveals 3,384 real-world GDPR consent violations across 94.1\% of analyzed forms. 

\section*{Acknowledgment}
We sincerely appreciate our shepherd and all anonymous reviewers for their insightful and valuable feedback. 
We thank Tamjid Al Rahat for feedback, and Bright Data for providing us with access to their proxies.
The work was supported in part by the NSF grants 2323105, 2317184, 2532587, Keysight Faculty Fellowship, 
and NSERC Discovery Grant RGPIN-2018-05931. David Lie is supported by Tier 1 Canada Research Chair CRC-2019-00242.
Wenjun Qiu is partially supported by an SRI fellowship.

\bibliographystyle{IEEEtran}
\bibliography{refs}

\normalsize
\appendices

\section{Web Form Extraction}
\label{appendix:wfe}

For visual processing, we capture full-page screenshots of the rendered web page. 
Since VLMs often struggle with extremely tall images because of resolution limitations and reduced spatial focus~\cite{cai2024spatialbot}. 
To mitigate this, we partition each screenshot into horizontal segments of 720 pixels in height, which approximates the typical viewport of standard browsers and aligns with how users visually perceive web content. 
This segmentation preserves visual clarity and improves the accuracy of form element identification. 
Each segment is independently analyzed by the VLM to extract form elements and consecutive segments containing web forms are merged into unified regions. 

\begin{algorithm}[ht]
    \caption{ConstructForm$(H, V) \rightarrow F$}
    \label{alg:element_matcher}
    \begin{algorithmic}[1]
        \State \textbf{Input:} $H, V \leftarrow$ HTML and Visual Element.
        \State \textbf{Initialize:} $\theta \leftarrow $ Threshold, $\alpha, \beta \leftarrow $ Text, Type Weight.
        \State $M \gets \emptyset$ \Comment{Matched pairs}.
        
        \For{$h \in H$}
            \State $best\_v \gets \text{None}$, $best\_score \gets \theta$\;
            
            \For{$v \in V$ not matched}
                \State $s \gets \alpha \cdot \text{TypeMatch}(h,v) + \beta \cdot \text{TextMatch}(h,v)$\;
                \If{$s > best\_score$}
                  \State $best\_v \gets v$, $best\_score \gets s$\;
                \EndIf
            \EndFor

          \If{$best\_v \neq \text{None}$}
            \State $M \gets M \cup \{(h, best\_v)\}$\;
          \EndIf
        \EndFor

        \State $U_H \gets H \setminus \{h \mid (h, -) \in M\}$ \Comment{ Unmatched HTML}
        \State $U_V \gets V \setminus \{v \mid (-, v) \in M\}$ \Comment{ Unmatched Visual}
        \State $T \gets \text{ExtractStaticText}(V)$ \Comment{Visual-only context}
        \State $F \gets \text{Merge}(M, U_H, U_V, T)$ \Comment{Final form}
        \State \Return{$F$}
    \end{algorithmic}
\end{algorithm}

The web form matching process is detailed in Algorithm~\ref{alg:element_matcher}.
Specifically, \textit{ConstructForm} takes a set of HTML elements $H$ as input and and visual elements $V$. 
Then it iterates over each HTML element $h \in H$ and identifies the best-matching visual element $v\in V$ that has not already been matched. 
For each candidate pair $(h,v)$, our algorithm computes a similarity score using a weighted combination of type compatibility and text similarity, where textual attributes include labels, placeholders, or values.  
After all HTML elements are processed, the algorithm collects unmatched HTML and visual elements as $U_H$ and $U_V$, respectively. 
To recover additional contextual information, it extracts standalone visual-only text segments (e.g., section headers or descriptive phrases not linked to form inputs) from $U_V$. 
Finally, all matched pairs, unmatched elements, and extracted textual context are merged into a unified form representation $F$.



\section{Used Prompts}
\label{appendix:prompt}

Here we list the prompts that we used to query LLM in Section~\ref{sec:method}.
\begin{figure}
    \centering
    \includegraphics[width=0.43\textwidth]{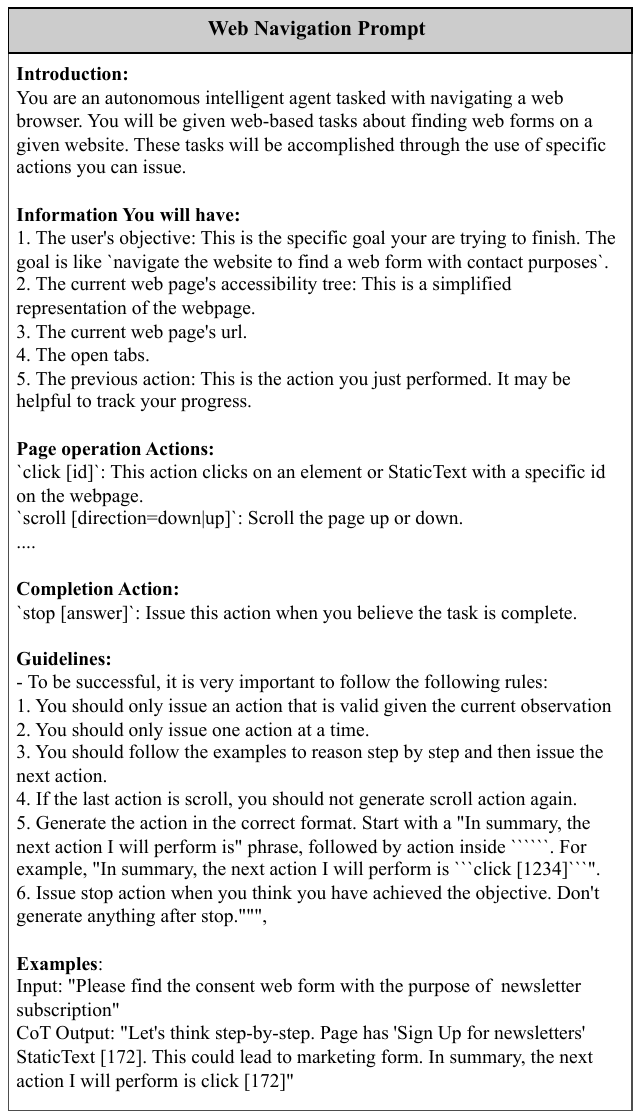}
    \caption{Prompt Template for Web Navigation.}
    \label{fig:webnav_prompt}
\end{figure}

\begin{figure}[h]
    \centering
    \includegraphics[width=0.46\textwidth]{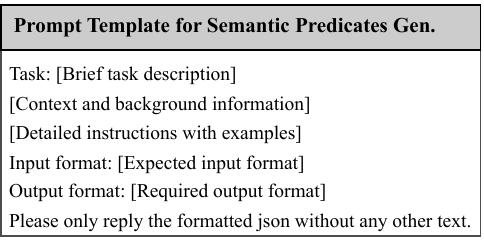}
    \caption{Prompt Template for Semantic Predicates Generation.}
    \label{fig:prompt_template}
\end{figure}

\begin{figure}
    \centering
\includegraphics[width=0.75\linewidth]{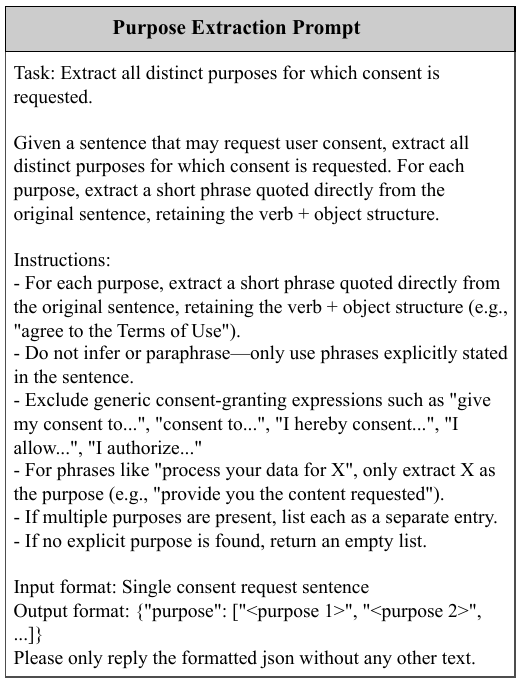}
    \caption{Purpose Extraction for Those Using Consent.}
    \label{fig:purpose_extraction_prompt}
\end{figure}

\begin{figure}[h]
    \centering
    \includegraphics[width=0.46\textwidth]{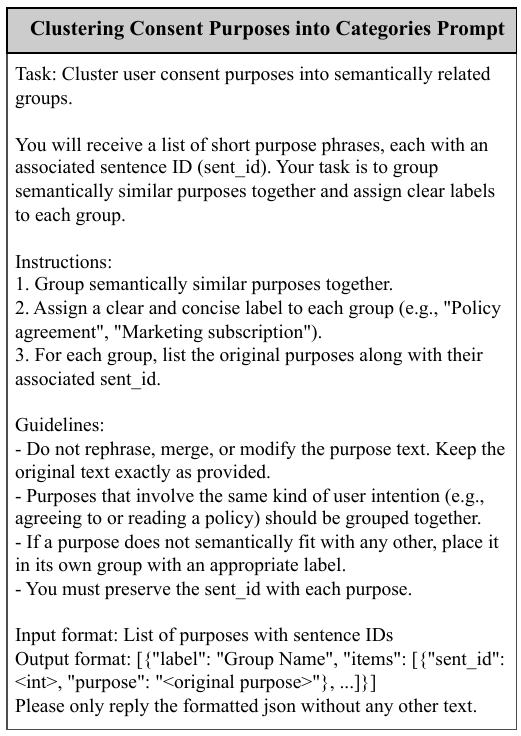}
    \caption{The used prompts for clustering purposes into different categories, where the results are directly mapped to generate semantic facts.}
    \label{fig:prompt_example}
\end{figure}

\section{Additional Figures}
\label{appendix:figure}

In this section, we list the additional figures mentioned in Section~\ref{sec: rq1}

\begin{figure}
    \centering
\includegraphics[width=0.75\linewidth]{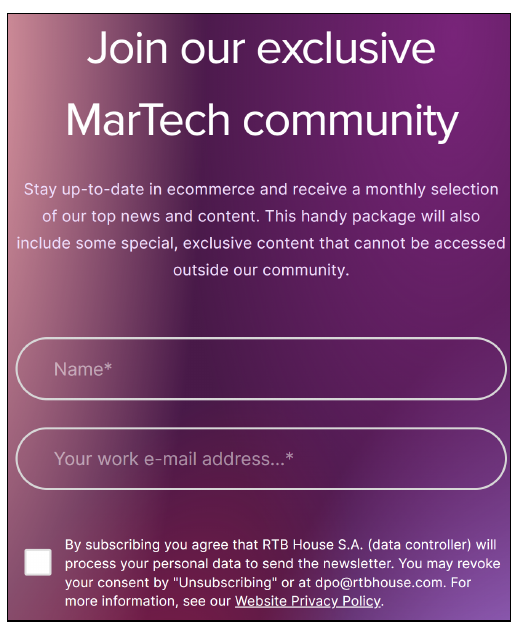}
    \caption{Real world example of a separate consent violation false positive of \tool. The form is extracted from \url{https://www.rtbhouse.com/newsletter}}
    \label{fig:sc_fp_case}
\end{figure}

%
\clearpage
\section{Meta-Review}
The following meta-review was prepared by the program committee for the 2026 IEEE Symposium on Security and
Privacy (S\&P) as part of the review process, as detailed in the call for papers.

\subsection{Summary}
The paper presents Cosmic, an automated framework that detects user's consent related GDPR violations in websites. To perform the compliance check, it performs the following tasks: 1) Cosmic utilizes LLM and LLM agent to locate the privacy policy and the corresponding web forms in a given website, and creates a logical representation of the web form using a specialized DSL, 2) it converts GDPR requirements into Datalog rules, and 3) it compares the Datalog rules against the DSL representation to identify potential GDPR violations. Finally, Cosmic is applied to 3598 forms collected from 5823 websites, and it identifies violations in 3384 (94.1\%) forms.

\subsection{Scientific Contributions}
\begin{itemize}
    \item Creates a New Tool to Enable Future Science
    \item Provides a Valuable Step Forward in an Established Field
\end{itemize}

\subsection{Reasons for Acceptance}
\begin{enumerate}
    \item \tool is an end-to-end framework for performing GDPR compliance checks. It is potentially extendable to other regulations.
    \item This paper provides a valuable step forward in automating the task of privacy compliance checks.
\end{enumerate}

\subsection{Noteworthy Concerns}
The underlying causes or trends of the potential compliance violations in the websites remain unexplored in the paper.

\section{Response to Meta Review}
We thank the anonymous reviewers for their insightful comments and the shepherd for the meta-review. We acknowledge the noteworthy concern raised in the meta-review. Our work primarily focuses on developing an automated system to detect GDPR consent violations. While understanding the underlying causes of these violations is indeed important, it requires user or developer studies, which we consider a valuable direction for future research.

\end{document}